\documentclass[fleqn,usenatbib,useAMS]{mnras}

\usepackage{graphicx}	
\usepackage{amsmath}	
\usepackage{amssymb}	

\usepackage[T1]{fontenc}
\usepackage{ae,aecompl}

\usepackage{newtxtext,newtxmath}

\usepackage{xcolor}

\newcommand{\lw}[1]{\textcolor{black}{#1}}
\newcommand{\Rsun}{\ensuremath{\rm{R}_\odot}}
\newcommand{\Lsun}{\ensuremath{\rm{L}_\odot}}

\newcommand{\Msun}{\ensuremath{\rm{M}_\odot}}

\DeclareRobustCommand{\DE}[3]{#3}

\title[Active dC stars with short orbital periods]{Carbon-enhanced stars with short orbital and spin periods}

\author[L.  J.  Whitehouse et al.]{L.  J.  Whitehouse,$^{1,2}$\thanks{E-mail: lewis.whitehouse.16@ucl.ac.uk}
J.  Farihi,$^{1}$
I.  D.  Howarth,$^{1}$
S.  Mancino,$^{2,3,4}$
N.  Walters,$^{1}$
A.  Swan,$^{1}$
\newauthor
T.  G.  Wilson,$^{1}$
J.  Guo$^{1}$
\\
$^{1}$Department of Physics and Astronomy, University College London, Gower Street, London WC1E 6BT, UK\\
$^{2}$European Southern Observatory, Karl-Schwarzschild-Strasse 2, 85748 Garching, Germany\\
$^{3}$Technical University of Munich Cluster of Excellence Universe, Boltzmannstr.  2 D-85748 Garching, Germany\\
$^{4}$Ludwig Maximilian University of M\"unchen, Geschwister-Scholl-Platz 1, 80539 M\"unchen, Germany
}

\date{Accepted XXX.  Received YYY; in original form ZZZ}

\begin{document}



\maketitle

\begin{abstract}
Many characteristics of dwarf carbon stars are broadly consistent with a binary origin, including mass transfer from an evolved companion.  While the population overall appears to have old-disc or halo kinematics, roughly 2\,per cent of these stars exhibit H$\upalpha$ emission, which in low-mass main-sequence stars is generally associated with rotation and relative youth.  Its presence in an older population therefore suggests either irradiation or spin-up.  This study presents time-series analyses of photometric and radial-velocity data for seven dwarf carbon stars with H$\upalpha$ emission.  All are shown to have photometric periods in the range 0.2--5.2\,d, and orbital periods of similar length, consistent with tidal synchronisation.  It is hypothesised that dwarf carbon stars with emission lines are the result of close-binary evolution, indicating that low-mass, metal-weak or metal-poor stars can accrete substantial material prior to entering a common-envelope phase.
\end{abstract}




\section{Introduction}

Binary-star evolution is responsible for a wide array of astrophysical phenomena, including blue stragglers, Type-Ia supernovae, and gravitational waves \citep{Blue_S,Webbink84,Grav_wav}.  Interactions between binary components may result in deviations from the course of typical single-star evolution, through mass transfer or mergers, leading to changes in stellar structure and atmospheric chemistry \citep{Hurley02,Merg}.  A prime example of how binary interactions can influence stellar evolution is given by dwarf carbon (dC) stars.  These low-mass, main-sequence stars exhibit spectra analogous to those of evolved carbon stars found at the tip of the asymptotic giant branch (AGB; \citealt{Dahn77}), but in advance of dredge-up processes being able to modify surface chemistry.

The peculiar atmospheric chemistry (C/O$\,>1$) in the prototype dC star, G77-61, is thought to originate through mass transfer from an evolved, higher-mass companion \citep{Dearborn86}.  In this picture, as the former primary star ascends the AGB, CNO-processed material is dredged to its surface before stellar winds liberate the consequently carbon-enhanced outer layers \citep{Iben}.  Enriched material is then transferred to the main-sequence companion, polluting its atmosphere and creating a dC star, while the original primary becomes a white dwarf.  The exact physical means of mass transfer in these systems remain unknown, but possibilities include classical processes of stellar-wind capture and Roche-lobe overflow, and more-recent models such as wind--Roche-lobe overflow \citep{Abate13}.  Regardless of the detailed mechanisms, metal-poor stars are significantly more susceptible to atmospheric pollution through external carbon enrichment processes \citep{Kool95}.

There is little optical evidence for white-dwarf companions to dC stars, where only $\lesssim1$\,per cent of systems exhibit flux consistent with a white dwarf at blue wavelengths (e.g.\ \citealt{Heber93,Liebert1994,Green2013}).  Nonetheless, recent observational evidence supports a binary formation channel for dC stars.  A radial-velocity study of 28 dC stars, each with typically 3--5 measurements obtained over several years, suggests a dC binary fraction of at least 75\,per~cent \citep{Whitehouse18}.  A larger survey of 240 dC stars, at a lower resolution, also found a high binary fraction, possibly as large as 95\,per cent, with a mean orbital period on the order of 1\,yr \citep{Roulston19}.

The first four, well-characterised dC binaries all have orbital periods on the order of years, as established by radial-velocity or astrometric monitoring \citep{Dearborn86,Harris18}: G77-61 (0.67\,yr), LSPM\,0742+4659 (1.23\,yr), LP\,255-12 (3.21\,yr), and LP\,758-43 (11.4\,yr).  These systems are all single-lined spectroscopic binaries or their astrometric equivalent, with neither optical nor ultraviolet indications of flux from their suspected evolved companions.  

The existence of long orbital periods ($\ga100$\,d) among dC stars is commensurate with better-studied classes of carbon-enhanced binary systems, such as CH, Ba, and carbon-enhanced metal-poor (CEMP) stars, which are all thought to follow analogous evolutionary pathways and which have orbital periods of hundreds to thousands of days \citep{CH,Ba,Cemps}.  In contrast, short-period binaries are not necessarily expected, as Roche-lobe overflow tends to be unstable in the case of mass transfer from a higher- to a lower-mass star, and the sole population-synthesis study for dC stars verifies that short-period systems fail to result if accretion is inefficient during a common-envelope phase \citep{Kool95}.  

However, in the same year that three of the long-period orbits were published, the dC star SDSS\,J125017.90$+$252427.6 (hereafter J1250) was reported to have a photometric period of 2.9\,d, and radial-velocity measurements folded on this period appear to provide a good orbital solution \citep{Margon18}.  This system has two notable features: photometric variability on a period potentially identical to the orbital period, and prominent emission in the Balmer lines and in Ca\,{\sc ii} H \& K, all of which are photospheric \citep{Margon18}.  Such emission features are rare among dC stars, with only a handful known among the $\sim$1200 confirmed and suspected candidates in the Sloan Digital Sky Survey \citep{Green2013}.  Furthermore, a {\em Chandra} study of six emission-line dC stars detected each of them in X-rays -- a clear indication of stellar activity \citep{Green19}.  Such activity was not anticipated for dC stars, as they are expected to be drawn from populations of older, metal-weak disc stars as well as truly metal-poor halo stars (including the prototype G77-61 with [Fe/H] $= -4$; \citealt{Plez05}) on the bases of their kinematics \citep{Harris98,Plant16,Farihi18} and population-synthesis modelling \citep{Kool95}.

Motivated by these findings, this study investigates the possibility that emission lines observed in some dC stars arise because of unexpectedly rapid rotation, associated with tidal synchronisation in short-period binaries, or with spin-up due to mass transfer.  This paper presents the results of radial-velocity and photometric time-series analyses conducted for seven dC stars with H$\upalpha$ emission, and finds that six stars in the sample are binaries with orbital periods in the range 0.2--4.4\,d.  All seven sources show photometric variability on periods that are either orbital, or are similarly short.  The radial-velocity and photometric observations are summarised in Section~2, the methods and period analyses are discussed in Section~3, the results for each individual target are presented in Section~4, and a discussion follows in Section~5.

\section{Observations and data reduction}

\subsection{Target selection}

The literature was searched to identify dC stars known to exhibit H$\upalpha$ emission (e.g.\ \citealt{Lowrance03,Green2013}), with six selected for further study based on brightness, visibility, as well as proper motion consistent with the main sequence, and are listed in Table~\ref{tab:Targets}.  Four have been monitored previously for radial-velocity variations over a several-year baseline, with a median of five observing epochs per target (J1015, CLS\,29, SBSS\,1310, CBS\,311; \citealt{Whitehouse18}).  Furthermore, the emission-line dC star J1250 is included in this analysis, though no new data have been obtained for this target.

\begin{table}
\centering
\caption{Observed dC stars with H$\upalpha$ emission.}
\label{tab:Targets}
\begin{tabular}{lcccc}

\hline

Target 			&\multicolumn{2}{c}{J2000 Coordinates}	&SDSS $r$	&Ref.\\
	 			&($^{\rm h\,m\,s}$)	&($\degr\,'\,''$ ) 		&(AB\,mag)	&\\
\hline

SDSS\,J084259 	&08 42 59.8 		&+22 57 29 		&17.1		&1\\
SDSS\,J090128 	&09 01 28.3 		&+32 38 33 		&17.1		&1\\
SDSS\,J101548 	&10 15 48.9 		&+09 46 49 		&16.9		&2\\
CLS\,29  			&10 40 06.4  		&+35 48 02 		&14.9  		&3\\
SBSS\,1310+561 	&13 12 42.4 		&+55 55 54 		&14.3		&4\\
CBS\,311 			&15 19 06.0 		&+50 07 02 		&17.4		&5\\

\hline

\end{tabular}
\flushleft
{\em Notes}.  The final column lists either the discovery or the first reference to illuminate the nature of the source: (1) \citealt{Green2013}; (2) \citealt{Whitehouse18}; (3) \citealt{Mould85}; (4) \citealt{Rossi2011}; (5) \citealt{Liebert1994}.  
\end{table}

\subsection{Radial velocities}

Radial-velocity monitoring was conducted using the ISIS spectrograph on the 4.2-m William Herschel Telescope at the Roque de Los Muchachos Observatory.  Some of the observations are reported in a previous paper \citep{Whitehouse18}, and further observing runs have since been carried out in 2018 Aug, 2019 Feb and Apr, and 2020 Feb.  All observations prior to 2020 were carried out using only the blue arm of the spectrograph, without any dichroic, in combination with the R1200B grating.  An EEV12 detector was employed without binning, providing wavelength coverage of approximately 5000--6000\,\AA\ and a resolving power $R\approx6700$ at the central wavelength (1\,arcsec slit).  The wavelength range encompasses many strong atomic and molecular lines, including the Mg\,{\sc i}\,b triplet, Na\,{\sc i}~D doublet, and two prominent C$_2$ Swan bands (see Table~\ref{tab:Wavelengths}).  

In 2020 Feb, both the blue and red arms of the ISIS spectrograph were used, with the GG495 dichroic.  The blue arm was equipped with the R600B grating to provide wavelength coverage of 3510--5340\,\AA, achieving a resolving power $R\approx2700$ at the central wavelength.  These data were obtained using the EEV12 detector without binning.  The red arm of the spectrograph was employed with the R1200R grating, covering a wavelength range of 5770--6770\,\AA, where a resolving power of $R\approx7400$ was achieved.  These data were obtained using the RED+ detector with no binning.  Together, these more recent data cover H$\upalpha$ and the higher Balmer lines.

Beginning in 2019, each star was observed a number of times (typically two to four) during each night to probe for short-term changes in radial velocity.  Each observation consisted of at least three separate exposures of the science target, with wavelength-calibration frames taken immediately before and after.  Individual exposure times for the stars varied between 180 and 600\,s typically, for the brighter and fainter sources, respectively.  Standard stars and other calibration data were taken at least once per night.

All science exposures were bias subtracted, flat fielded, and extracted using standard methods in {\sc iraf}.  For extraction, the 2-dimensional spectral trace was established using the standard star observation with the highest signal-to-noise (S/N) of the night and was re-centred for the extraction of each target spectrum.  Individual exposures were wavelength calibrated using the appropriate arc frames for each target.  Finally, all consecutive target exposures were combined using inverse variance weighting, to create a single spectrum with S/N $\ga10$ for analysis.

\subsection{Photometry}

Time-series photometry used in this study was sourced primarily from the public data releases of the Palomar Transient Factory (PTF), intermediate-Palomar Transient Factory, and Zwicky Transient Facility (ZTF) surveys \citep{Law2009,Kulkarni2013,Bellm2014}.  The primary science goals of these projects are to monitor the sky for transient events, using the Palomar 48-in Oschin Schmidt Telescope with wide-field imagers, surveying the entire available northern sky roughly every three days.  The detection limit of the survey is $R \approx 20.6\,$mag \citep{Rau2009}, and all data are fully reduced and calibrated by the dedicated facility pipelines \citep{Ofek2012,Masci2019}.

Data were retrieved for J0901, J1015, and CBS\,311 by querying the PTF~DR3 and ZTF~DR5 databases using their \textit{Gaia} eDR3 astrometry \citep{eDR3}.  For each target, a cone search was made using a 15-arcsec radius centred on the 2015.5 \textit{Gaia} positions.  Only matches within 0.5\,arcsec of the expected position were selected for analysis; this is well within the typical diameter of the point spread function of the surveys.  All photometric data with poor quality flags were rejected.

Photometric monitoring of the dC star J0842 was conducted by \textit{Kepler} during \textit{K2} campaign 18, with continuous measurements for two months at a cadence of 30\,min \citep{Howell14}.  Other sources of photometry were not used for this target, as the \textit{Kepler} data are of superior quality.

Finally, CLS\,29 and SBSS\,1310 were observed with the {\em Transiting Exoplanet Survey Satellite} ({\em TESS}; \citealt{TESS}), and are sufficiently bright and isolated that useful data were obtained.  CLS\,29 was observed in Sector 22, and SBSS\,1310 in Sectors 15, 16, and~22 (with an interval of 139~days between Sectors 16 and~22).  The duration of each {\em TESS} sector is approximately 25\,d, with photometric observations taken every 30\,min.

\section{Methods and period analysis}

\begin{table}
\centering
\caption{Wavelength regions and atomic transitions used for spectral cross-correlation to determine absolute velocities.}
\label{tab:Wavelengths}
\begin{tabular}{cl} 

\hline

Wavelengths		&Important lines \\
(\AA)				& \\

\hline

4092 -- 4112  	&H$\updelta$ 				\\
4330 -- 4350  	&H$\upgamma$ 			\\
4851 -- 4871  	&H$\upbeta$  				\\
5163 -- 5187 	&\ion{Mg}{i} b 				\\
5197 -- 5215 	&\ion{Fe}{i} / \ion{Cr}{i} blends \\
5258 -- 5273	&\ion{Fe}{ii} 				\\
5325 -- 5331 	&\ion{Fe}{i} 				\\
5368 -- 5376 	&\ion{Fe}{i} 				\\
5863 -- 5924 	&\ion{Na}{i} D 				\\
6553 -- 6573  	&H$\upalpha$ 				\\

\hline

\end{tabular}

\end{table}

\subsection{Radial-velocity cross-correlation}

The most precise measurements of radial-velocity changes are achieved by cross-correlating individual spectra of a given target against its highest S/N counterpart.  Such measurements benefit from the strong molecular carbon absorption features (e.g.\ C$_2$ Swan bands) in dC stars.  To achieve the best cross-correlation possible, all spectra were first continuum-normalised by fitting a cubic spline.  The cross-correlation was performed using the \textsc{iraf fxcor} package \citep{TD1979}.  This cross-correlation method reduces the velocity error by around 50\,per cent compared to a determination of absolute velocity using the few available (weak) atomic lines.  These more precise but relative velocity measurements are then used to determine the orbital parameters for each target.

To establish an absolute velocity scale for these data, the highest S/N spectrum of each target was cross-correlated against a continuum-normalised synthetic spectrum template.  These spectra were synthesised using {\sc autokur}, which is a wrapper for the Kurucz spectral synthesis codes \citep{Kurucz2005}, with the underlying atmospheric structures based on {\sc marcs} models \citep{Marcs08}.  Stellar parameters were chosen to be representative of low-mass main-sequence stars (i.e.\ appropriate for late-K to early-M dwarfs), and to reflect the older kinematical ages of the dC population as a whole.  Templates were generated for $T_{\rm eff}=3500$--4500\,K in steps of 250\,K, and metallicities [Fe/H] from $-2$ to $0$ in integer steps, with the surface gravity and microturbulence kept fixed at $\log\,[g\,{\rm (cm\,s}^{-2})] = 5.0$ and $v_{\rm t} = 2.0\,$km\,s$^{-1}$, respectively.  The synthetic spectra were generated at resolving powers of $R=6700$ and 7900 to match the instrumental resolution achieved under standard and best observing conditions at the telescope, respectively.  

Because stellar-atmosphere models are not currently available for cool, carbon-rich dwarf stars, the synthetic spectra were generated for stars with solar carbon abundances, at gravities and effective temperatures similar to those expected for a typical dC star.  The C$_2$ Swan bands, which contain myriad individual transitions and are useful for relative velocity measurements, are not present in the solar-abundance templates, and would lead to increased measurement errors in velocity if used as the primary source of cross-correlation.  Therefore, to minimise the measurement errors, restricted wavelength regions of the synthetic templates -- those containing the strongest atomic lines -- were selected for cross-correlation.  Using these restricted regions significantly improved the quality of each cross-correlation (albeit while still not as precise as the relative velocity measurements), by mitigating spectral mismatches.  The wavelength regions used for the synthetic templates are listed in Table~\ref{tab:Wavelengths}.

To determine the best synthetic template to measure absolute velocities, each synthetic template was cross-correlated against the highest S/N spectrum of each observed dC star.  The template that achieved the strongest cross-correlation function was therefore deemed as the most suitable.  In general, templates at cooler temperatures achieved a stronger cross-correlation function than templates at higher temperatures; those with $T_{\rm eff}\leq4000$\,K consistently outperformed their warmer counterparts across the whole metallicity range.  Furthermore, templates with metallicity [Fe/H] $= 0$ typically performed worse when compared to lower metallicity templates.  Overall, the resolving power of the templates exhibited no performance trend in cross-correlations.  The stellar parameters of the best performing synthetic spectral template were [Fe/H] $= -1$, $T_{\rm eff}=3750$\,K, and $R=6700$.  This template was therefore adopted to measure absolute velocities, as it consistently possessed strong cross-correlation metrics with high S/N observed spectra.

\subsection{Radial-velocity period determinations}
\label{SecRVPD}

The {\sc joker} is software designed to determine orbital periods from radial-velocity data, under the assumption that all variability is the product of two-body gravitational interactions \citep{Price2017,Joker}.  For a given number of samples drawn from \lw{a} prior probability distribution function (PDF), it uses the method of rejection sampling to eliminate low-likelihood solutions, generating a random value for each posterior sample, $r_{\rm j}$, between 0 and the maximum likelihood in the posterior distribution, $L_{\rm max}$.  If the likelihood of a specific sample, $L_{\rm j}$, does not meet the condition that $r_{\rm j} < {L}_{\rm j}$, then the sample is rejected.  This approach guarantees at least one surviving sample and is particularly useful where radial-velocity data are sparse.

Prior PDFs were selected for orbital eccentricity, velocity semi-amplitude, and systemic velocity as follows.  Only circular orbits were considered, both for simplicity and because the available radial-velocity sampling is insufficient to robustly characterise eccentric orbits.  Should a short-period solution result for any given system, then the circular orbit assumption is retrospectively well justified for post-common-envelope binaries \citep{Pac76,Pod01}.  Velocity semi-amplitudes were sampled from a Gaussian PDF: for each target, the mean was set to zero, with the standard deviation set to 110\,km\,s$^{-1}$ (taking the absolute value).  The adopted standard deviation is based on the expected velocity semi-amplitude for a binary with a total mass of 1.1\Msun, a 5\,d period, and 60\degr inclination.  The width of the prior PDF on velocity semi-amplitude is intended to avoid biasing the posterior distributions.  Finally, systemic velocities were sampled from a Gaussian distribution also centred at zero, with a standard deviation of 100\,km\,s$^{-1}$.  Again, this prior PDF is uninformative and should not bias against any actual velocities.  

Because the \textsc{joker} analysis is performed on the more precise, relative radial-velocity data, the systemic velocity returned from the analysis is offset from the true value.  This was corrected by applying the absolute velocity offsets as determined by the synthetic spectral templates, while propagating the errors in the offset measurement (which are assumed to be Gaussian) with the systemic velocity posterior distribution.

To ensure that an adequate number of posterior samples survive the \textsc{joker} analysis, $10^9$ prior samples are generated for each target, with the maximum posterior sample size set to 8192.  However, the number of posterior samples removed during the rejection sampling algorithm in the {\sc joker} is a function of the maximum likelihood of those posterior samples.  Therefore, for targets possessing more constraining data, fewer posterior samples survive.

In the scenario where few samples survive rejection sampling, the posterior distribution returned from the \textsc{joker} is typically unimodal with fewer than 100 samples.  Owing to the small number of surviving samples, in order to explore the posterior distribution fully, these surviving samples are used as prior samples to initialise a Markov Chain Monte Carlo (MCMC) simulation.  The \textsc{pymc3 no-u-turn} sampler was used with four chains and a tuning length of 1000, yielding a total of 2500 posterior samples per chain \citep{pymc3,NUTS}.  This further analysis with MCMC was necessary for the dC stars CLS\,29 and SBSS\,1310, as only 10 and 74 samples survived rejection sampling, respectively.  All other targets possess sufficiently large posterior distributions from the \textsc{joker} such that this additional analysis was obviated.

\begin{figure}
\includegraphics[width=1.0\columnwidth]{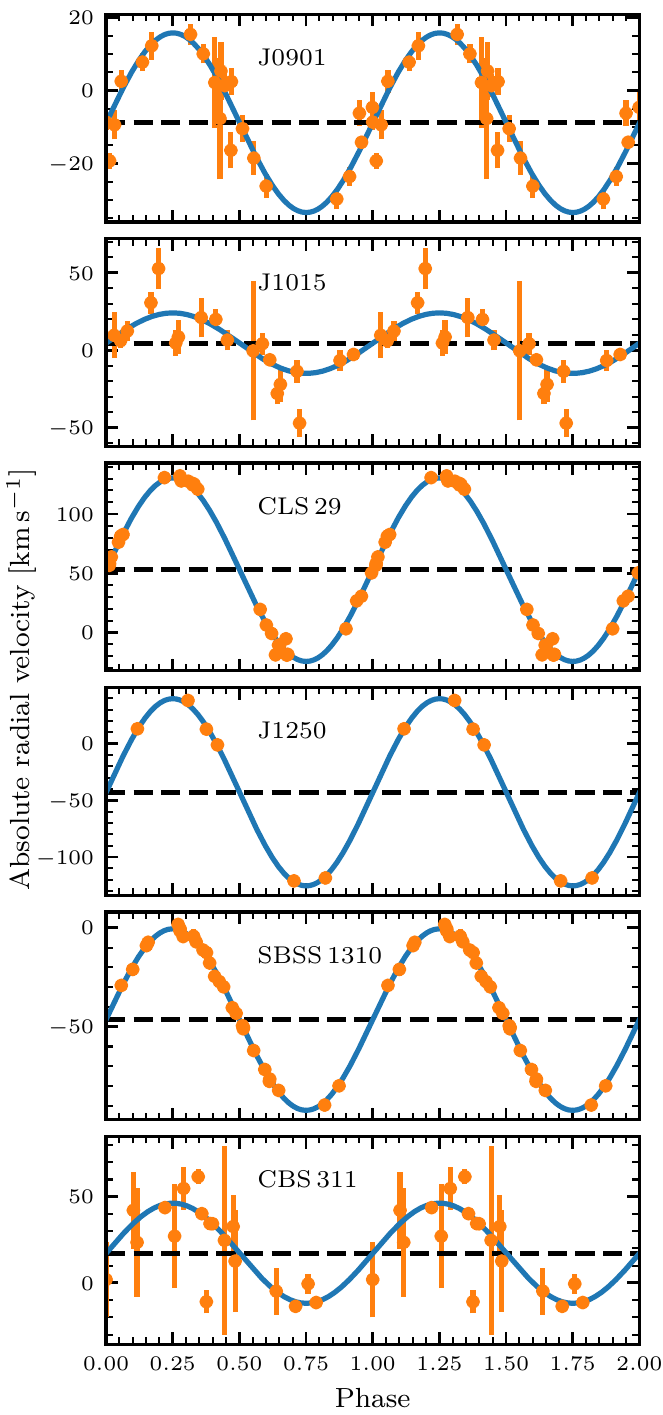}
\vskip -5 pt
\caption{Radial-velocity curves corresponding to the median, posterior parameter values detailed in Section~\ref{SecRVPD}.  For each target, the radial-velocity data and corresponding measurement errors are shown in orange, phase-folded and fitted in blue with the determined orbital parameters.  The dashed black line corresponds to the median of the fitted curve, representing the systemic velocity.  The orbital period and reduced $\upchi^2$ are given in Table~\ref{tab:Summary}.}
\label{fig:RV}
\end{figure}

The orbital parameters for each target are quoted as the median of the final posterior distribution (whether through the \textsc{joker} or MCMC) and presented in Table~\ref{tab:Summary}.  The median was selected due to the non-Gaussian nature of some of the distributions.  The uncertainties in the orbital parameters for each target are quoted as the 16\,per cent and 84\,per cent confidence limit in the posterior distribution.  This analysis was completed for the six targets discussed in this paper, as well as J1250.  The radial-velocity curves corresponding to the median orbital parameters derived through this analysis are shown in Figure~\ref{fig:RV}, with the posterior PDFs of each target displayed in Figure~\ref{fig:all_corner}.

\begin{figure*}
\includegraphics[width=\textwidth]{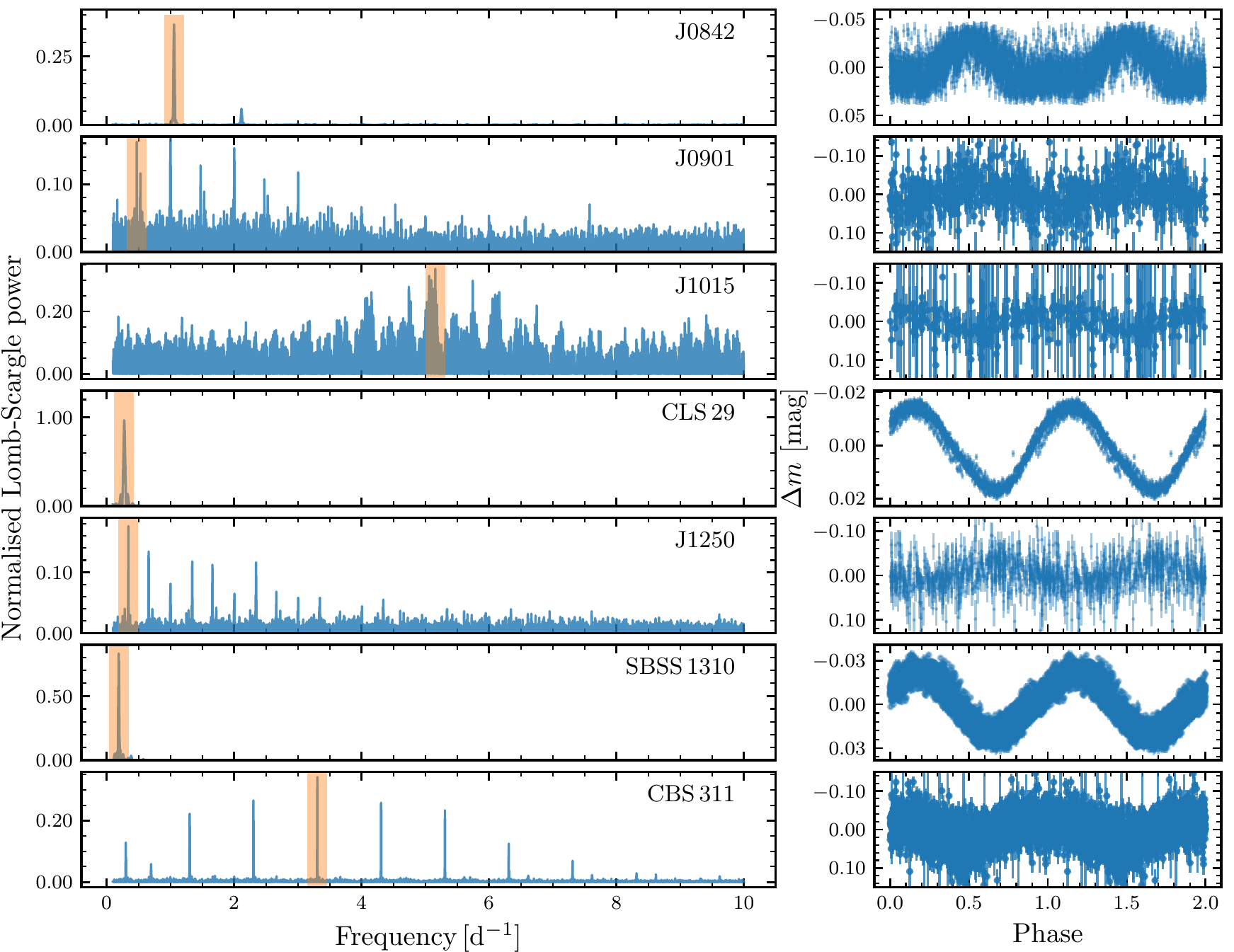}
\vskip -5 pt
\caption{\textit{Left:} Lomb-Scargle periodograms for the six dC stars studied here, in addition to J1250.  The most significant peak in the periodogram is highlighted with an orange vertical bar.  \textit{Right:} The phase-folded light-curves corresponding to the strongest periodogram peak for each target.  }
\label{fig:LC}
\end{figure*}

\subsection{Periodogram analysis of photometry}\label{P_analysis}

All photometric time-series were normalised to the median value for each target, then analysed using the {\sc astropy} implementation of the Lomb-Scargle (LS; \citealt{Lomb1976,Scargle1982}) periodogram, and the {\sc p4j} \citep{P4J} implementation of Multi-Harmonic Analysis of Variance (MHAOV; \citealt{Schwarz1996}).  The results of the LS analysis are presented in Figure~\ref{fig:LC}.

The LS periodogram is a method to derive the period of oscillation, for a recurrent signal, in unevenly sampled time-series data through the construction of a Fourier-like power spectrum \citep{Van2018}; it is almost equivalent to performing a least-square fit of a sine wave (cf.\ \citealt{Zech09}).  Such a periodogram can be relatively difficult to interpret, as the Fourier transform of any continuous signal in the data is convolved with the window function.  It can therefore be challenging to identify which periodogram peaks represent real signals in the time-series data, especially for relatively faint and modest S/N photometric sources such as dC stars \citep{Roulston2021}.  To break the degeneracy between the true and aliased peaks, an LS periodogram was computed for the window function for each target, and the window-function periodogram compared to the periodogram computed from the actual fluxes.

For an LS periodogram peak to be considered real, it must be sufficiently separate in frequency from a significant peak in the periodogram of the window function.  Quantitatively, a significant peak in the window function was defined as any peak with power above 20\,per cent of the strongest peak in the window function.  The minimum separation in frequency space between a window function peak and an LS peak was chosen to be 0.001\,d$^{-1}$.  Finally, a false-alarm probability was calculated by performing a bootstrap on the photometric data for the top five peaks that satisfied these conditions.  All peaks surviving these quality checks were then considered to be real periodic signals in the data.

To calculate the error in the photometric period corresponding to the best LS periodogram signal for a particular target, a random number of data points were dropped from the dataset and replaced by randomly sampling the original dataset.  This new dataset was then analysed as outlined above, and the process was repeated $10^3$ times to generate a distribution of most likely photometric periods, where the final error is quoted as the standard deviation of this distribution.  The LS periodograms for the seven targets and their most significant frequency peaks are shown in Figure \ref{fig:LC}.

\begin{table*}
\centering
\caption{Summary of period and close binary parameter estimates for dC stars, and possible sources of photometric variability.}
\label{tab:Summary}
\begin{tabular}{lcrrcccccrcccc} 
\hline

Star			&$P_{\rm rv}$	&$K_1$		&$i_{\rm mod}$ &$\upchi^{2}_{\rm red}$	&$P_{\rm phot}$	&$\Delta m$	&Distance	&$M_r$	&$m_{\rm NUV}$&\multicolumn{3}{c}{Variability?}\\	
			&(d)			&(km\,s$^{-1}$)	&($\degr$) 	&					&(d)				&(mag)		&(pc)	&(AB\,mag)&(AB\,mag)	&(T)&(I)&(R)\\

\hline	

J0842		&...			&$<13.8$		&$<6.1$		&...					&0.95			&$\pm0.014$	&680		&8.0		&22.0		&$-$	&$-$	&$+$\\
J0901		&2.12		&24.5		&14.3		&2.74				&2.12			&$\pm0.023$	&590		&8.2		&...			&$-$	&$-$	&$+$\\
J1015		&0.20		&18.5		&4.9			&2.19				&0.20			&$\pm0.022$	&480		&8.5		&18.0		&$-$	&$+$&$+$\\
CLS\,29		&3.15		&79.2		&65.3		&2.94				&3.63			&$\pm0.014$	&235		&8.0		&22.9		&$-$ &$-$ &$+$\\
J1250		&2.56		&86.1		&67.2		&0.47				&2.92			&$\pm0.023$	&280		&9.2		&$>22.0$		&$-$	&$-$	&$+$\\
SBSS\,1310	&4.43		&46.5		&36.7		&0.53				&5.26			&$\pm0.015$	&100		&9.2		&20.3		&$-$	&$-$	&$+$\\
CBS\,311		&0.19		&29.2		&7.6			&4.83				&0.30			&$\pm0.030$	&440		&9.2		&...			&$-$	&$+$&$+$\\
		
\hline
\end{tabular}
\flushleft	
{\em Notes.} Errors for the period determinations are non-Gaussian and discussed in detail for all stars in Section~4, with posterior distributions shown in the Appendix.  The third column lists a model binary inclination based on the adopted or suspected radial-velocity period and velocity semi-amplitude (or upper limit), under the assumptions of 0.4\,$\Msun$ for the dC star, and 0.6\,$\Msun$ for a white-dwarf companion.  The fourth column lists the reduced $\chi^{2}$ for each of the orbital solutions fitted in Figure~\ref{fig:RV}.  The final column gives a positive ($+$) or negative ($-$) indication as to whether the observed photometric variability amplitude could plausibly result from modulation due to tidal distortion (T), a heating effect from irradiation (I), or rotation (R).
\end{table*}

\subsection{Multi-harmonic analysis of photometry}

The MHAOV method was then used as an independent check on the LS periodograms.  In principle, this method differs from the LS analysis in terms of the model fitted to the data, and the statistic calculated from that fit.  Whereas the LS method assumes the data are well modelled by a sine wave, the MHAOV method instead uses orthogonal Szeg\"o polynomials.  It should be noted that the first harmonics of the Szeg\"o polynomials simplify to a sine wave, and thus in this work, both the LS and MHAOV method are fitting sine waves to the photometric data. 

The advantage of using MHAOV, therefore, comes from how the goodness of fit statistic is calculated \citep{Lachowicz2006}.  \lw{The statistic used in the LS method is simply $\upchi^2$, and depends on the variance of the data.  However, because the true data variance is not actually known, but rather estimated from the data, the goodness-of-fit statistic for the LS periodogram and variance estimations are not independent.}  The MHAOV method circumvents this problem by using the Fisher analysis-of-variance statistic instead of a $\upchi^2$, thus breaking the dependency between the statistic and the variance.  This has the benefit of suppressing the significance of periodogram peaks that are due to aliasing in the data; therefore, allowing for the determination of real peaks to be significantly easier than in the LS analysis.  Furthermore, the MHAOV analysis is capable of detecting periodic signals at lower S/N than the LS analysis, while also assigning real signals with a higher significance than the LS analysis \citep{Schwarz1996}.  Based on these statistical advantages, the MHAOV method was used in parallel to the LS method, and in particular to verify the five strongest peaks identified from the LS analysis for each target.  The MHAOV analysis agreed with the strongest peak identified through the LS analysis for all targets in this study and hence, only the LS periodograms are shown in Figure~\ref{fig:LC}.  All photometric periods quoted in this study are the most significant peak in both analyses.

\section{Results for individual stars}

The results of both the photometric and radial-velocity analyses are presented here in detail for each target, with a brief summary provided in Table \ref{tab:Summary}.  

\subsection{J0842}

\begin{figure*}
\includegraphics[width=\textwidth]{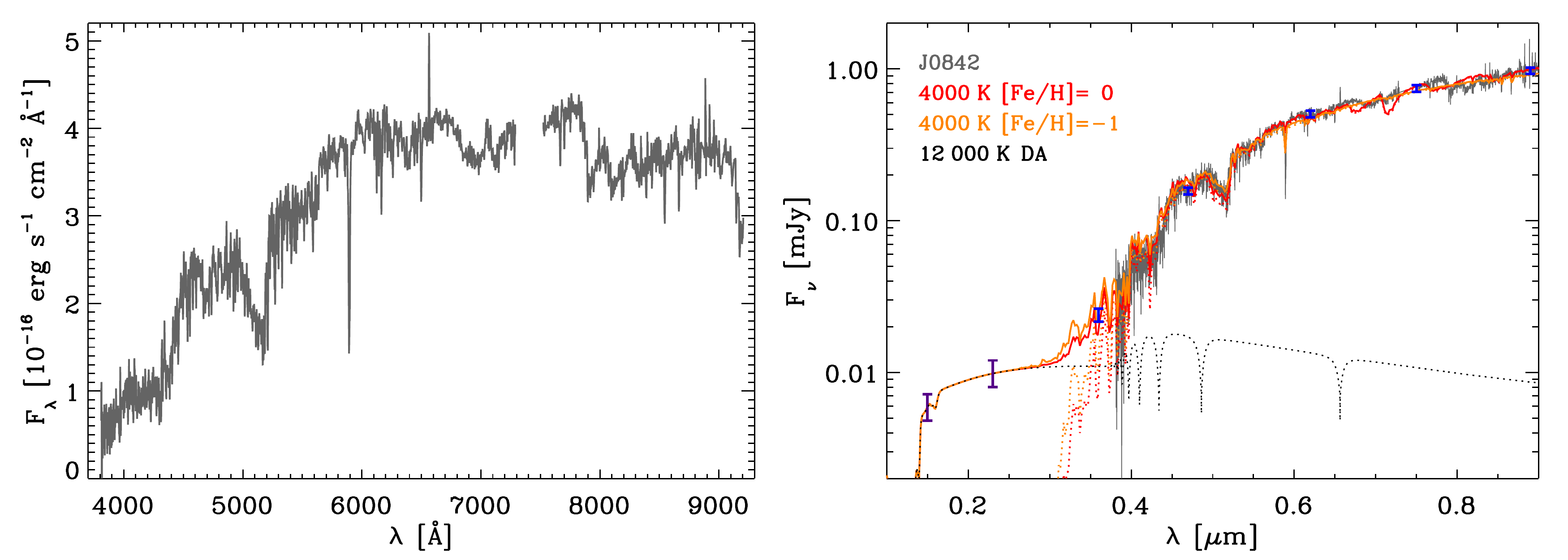}
\vskip -5 pt
\caption{{\em Left:} The spectrum of SDSS\,J084259.79+225729.8, which exhibits both carbon-normal and carbon-enhanced features.  \lw{This may suggest an intermediate C/O abundance, and this star may therefore be an example of an S-type dwarf \citep{VEck17}.}  The most prominent feature is the MgH band near 5100\,\AA, while other notable absorption lines include Na\,{\sc i} D, the K\,{\sc i} doublet near 7700\,\AA, and the Ca\,{\sc ii} red triplet.  The spectrum also shows a depression near 6800\,\AA \ due to CaH, which is a typical sign of metal poverty in cool dwarf stars.  {\em Right:} The ultraviolet through optical spectral energy distribution of J0842.  The SDSS spectrum is shown in solid grey, with photometric fluxes shown as blue and purple error bars.  Two scaled 4000\,K synthetic spectra are shown in dashed orange and red for different metallicities.  The addition of a 12\,000\,K, hydrogen-atmosphere white dwarf is shown in dashed black, and the resulting composite spectra are traced in solid orange and red.}
\label{fig:J0842}
\end{figure*}

The dC star J0842 has an unusual spectrum compared to its siblings, as is shown in Figure~\ref{fig:J0842}.  While the spectrum exhibits clear CN features in the red, below roughly 7000\,\AA \ the flux is broadly consistent with a late-K dwarf.  However, there are subtle but convincing signs of C$_2$ Swan bands in the blue, especially when compared side by side with an appropriate stellar template of solar abundance (e.g.\ the late sdK stars SDSS\,J110140.90+385230.2 or SDSS\,J142617.53+073749.2).  Also noteworthy is the CaH band -- a hallmark of metal-poor, cool subdwarfs \citep{Reid05,Lepine07}.  In addition to H$\upalpha$, close inspection also reveals emission in H$\upbeta$ and H$\upgamma$.  Without the clear CN features, the spectrum might easily be mistaken for that of a K~dwarf, although the emission lines would remain remarkable.

Interestingly, the source is detected with {\em Galex} in both the far- and near-ultraviolet bandpasses.  There are three and four separate detections respectively, with weighted averages of $m_{\rm FUV}=22.0$\,AB\,mag, $m_{\rm NUV}=21.4$\,AB\,mag.  It is possible that these ultraviolet detections are a result of strong cool-star line emission at these wavelengths, but there are few transitions that could be responsible.  In the near-ultraviolet {\em Galex} band, the Mg\,{\sc ii} 2800\,\AA \ doublet is a candidate, but this is highly unlikely.  It is near the red cutoff of the bandpass, and furthermore, the emission would have to be three orders of magnitude brighter than the stellar continuum.  The corresponding far-ultraviolet detection cannot be due to any common stellar emission features, as the bandpass cuts off before Ly$\upalpha$.  Therefore, the observed ultraviolet flux is almost certainly not intrinsic to the dC star.

The ultraviolet fluxes appear to be consistent with a white dwarf, and in Figure \ref{fig:J0842} the ultraviolet and optical photometry for J0842 is approximated using models.  Solar-abundance models \citep{Kurucz03} were used for the dC star, as there are currently no atmospheric models for carbon-enhanced cool dwarfs, while white-dwarf atmospheric models were taken from \citet{Koester10}.  A reasonable fit to the ultraviolet-optical photometry is achieved using a $T_{\rm eff}=12\,000$\,K, hydrogen-atmosphere white-dwarf model, together with a $T_{\rm eff}=4000$\,K main-sequence stellar template, where a modestly metal-poor model atmosphere provides a somewhat better fit than one with solar metallicity.

If the white dwarf is at the {\em Gaia} parallax distance of $d=680\pm50$\,pc, then, for the observed temperature and luminosity, evolutionary models\footnote{http://www.astro.umontreal.ca/~bergeron/CoolingModels} suggest a surface gravity close to a canonical value of $\log\,[g\,{\rm (cm\,s}^{-2})] = 8.0$, with a corresponding mass and radius of 0.61\,$\Msun$ and 0.013\,$\Rsun$, and cooling age of around 40\,Myr \citep{Fontaine01}.  It is noteworthy that the SDSS $u$-band photometry appears to be consistent with the composite stellar fluxes, and the {\em Gaia} distance indicates clearly that this is a main-sequence star with $M_r=8.0$\,AB\,mag.  

This target was observed during $K2$ campaign 18.  The LS periodogram analysis yields a clearly-defined photometric period of $0.945\pm0.001$\,d with a mean semi-amplitude of $\pm0.014$\,mag (Figure \ref{fig:LC}).  

Nine radial-velocity observations were obtained over three nights in 2019, with two to four measurements per night.  Additionally, five further radial-velocity measurements were taken in 2020 Feb.  Perhaps surprisingly, these data do not reveal any radial-velocity variations exceeding the error bars (typically $\pm$5\,km\,s$^{-1}$).  Thus, if J0842 is binary with the suspected white-dwarf companion described above, and its orbital period is similar to the observed photometric period of 0.95\,d, then the inclination must be $\le6.1\degr$ ($3\upsigma$) for masses of $M_{\rm dC}=0.4\,\Msun$ and $M_{\rm WD}=0.6\,\Msun$.

\subsection{J0901}

This source exhibits a dC optical spectrum with no evidence for a luminous companion, and is, therefore, a more typical example of its class, albeit with the exception of the H$\upalpha$ and likely H$\beta$ emission features.  It is detected in X-rays by {\em Chandra} \citep{Green19}, but {\em Galex} observations do not exist for this region of the sky.  This study is the first to address its potential binarity, and all radial-velocity observations reported are new.

The {\sc joker} radial-velocity analysis indicates that the orbital period distribution is approximately bimodal, with one larger peak at 2.11\,d and a second, smaller peak at around 2.09\,d.  The median of the posterior distribution yields a period of 2.12$^{+0.0}_{-0.03}$\,d and a velocity semi-amplitude of 24.49$^{+1.5}_{-1.6}$\,km\,s$^{-1}$.  A modest number of additional radial-velocity measurements at around a phase of 0.7 would likely precisely constrain the orbital period.  

The photometric analysis for J0901 was conducted on both ZTF and {\em TESS} data.  Because of the target faintness, the {\em TESS} photometry are not useful, but the ZTF data indicate a possible photometric period of 2.12\,d.  However, this photometric period remains relatively uncertain.

\subsection{J1015}

This source does not appear in any compendium of dC stars published to date but is flagged as having carbon lines by the SDSS spectroscopic reduction pipeline in Data Release 9.  The SDSS spectrum appears to be a composite of a dC star and a DA-type white dwarf that contributes continuum flux and broad absorption at H$\upgamma$, H$\updelta$, and H$\upepsilon$.  It is a binary suspect reported by \citet{Whitehouse18}, based on only two radial-velocity measurements which differ by nearly $7\upsigma$.

The white dwarf is detected clearly in {\em Galex} ultraviolet photometry, where the weighted average of two observations yield $m_{\rm FUV}=17.6$\,AB\,mag, $m_{\rm NUV}=18.0$\,AB\,mag.  The ultraviolet colour alone suggests an effective temperature between 20\,000 and 25\,000\,K for a typical white-dwarf surface gravity.  The inclusion of the SDSS $u$-band photometry and the $d\approx480$\,pc {\em Gaia} parallax distance then favours $T_{\rm eff}$ closer to, but not quite 25\,000\,K to best match the expected absolute ultraviolet magnitudes for a carbon--oxygen core white dwarf.  While J1015 is one of the six dC stars detected in X-rays by {\em Chandra} \citep{Green19}, the white dwarf cannot be the source of hard X-rays.  

Owing to the low S/N in the spectra of J1015, it was not immediately possible to cross-correlate each spectrum with the relative velocity template.  Therefore, each observed spectrum of J1015 was re-binned to lower resolution, with each bin consisting of three pixels.  These re-binned spectra were then cross-correlated against the relative velocity template to yield precise velocities for determining the orbital period, before being cross-correlated against the synthetic template, to determine absolute velocities.  

The \textsc{joker} radial-velocity analysis finds that the orbital period is strongly peaked at 0.20\,d.  There are a modest number of posterior samples discretely scattered at either side of this peak, though these samples are very few compared to the peak.  The velocity semi-amplitude of J1015 is measured to be $18.5^{+2.7}_{-2.5}$\,km\,s$^{-1}$, and thus the system inclination is likely to be low.

The photometric analysis reveals a strong frequency peak corresponding to a period of $0.196\pm0.003$\,d, where two smaller flanking peaks in the power spectrum are the result of windowing.  Relatively large photometric errors (averaging 0.07\,mag) for this source give rise to the broadened structures in the periodogram.  This photometric period is strongest in both the LS and MHAOV analyses, and the phase-folded light-curve at this period shows an amplitude near 2\,per cent.

\subsection{CLS\,29}

The optical spectrum of this dC star is relatively unremarkable, except for the presence of H$\upalpha$ and H$\upbeta$ emission.  The source is possibly marginally detected in the near-ultraviolet with \textit{Galex} at $m_{\rm NUV}=22.9\pm0.6$\,AB\,mag, but too uncertain to warrant further analysis.  The {\em TESS} data yield a clear peak in the periodogram, corresponding to a period of 3.62\,d.  The amplitude of the photometric variation is around 1.4\,per cent.  

This study is the first to report radial-velocity variations in CLS\,29.  The radial-velocity analysis is highly constraining, with only 1 in $10^8$ posterior samples surviving.  The MCMC analysis yields a bimodal distribution in the orbital parameters, where both peaks in the orbital period are within 1\,per cent of 3.14\,d.  Owing to incomplete phase coverage during the radial-velocity monitoring, the period cannot be uniquely determined.  Additional velocity measurements for phases $0.4-0.5$ should break the degeneracy between the two solutions.

\subsection{J1250}

This dC star is a known single-lined binary and the first shown to have a short-period orbit \citep{Margon18}, where both photometric and radial-velocity variations are observed.  It is included here for context and discussed later, as it belongs to the subgroup of stars exhibiting H$\upalpha$ emission lines, and is detected in X-rays \citep{Green19}.  There are {\em Galex} observations for this region of the sky, but no detection near the location of J1250 in either ultraviolet band, placing a rough upper limit at either wavelength of $m>22.0$\,AB\,mag.  For a white-dwarf companion at the {\em Gaia} parallax distance, this would imply a star no brighter than around $m=15$\,AB\,mag in the ultraviolet, and for a typical mass and radius this would translate to $T_{\rm eff}\la8000$\,K.

The publicly available photometric data from ZTF for J1250 were analysed using the LS and MHAOV analyses.  The photometric period of 2.92\,d identified by \cite{Margon18} was successfully recovered.  

While the radial-velocity measurements of J1250 taken by \cite{Margon18} do appear consistent with the photometric period, no dedicated radial-velocity analysis was performed in that study.  Here, the reported radial-velocity data are analysed using the \textsc{joker}, where the median posterior distribution for the orbital period is found to be $2.56^{+0.13}_{-0.11}$\,d, and is notably 12\,per cent shorter than the reported photometric period.

\subsection{SBSS\,1310}
\label{subsec:sbssp}

This is one of the nearest and brightest dC stars known, and among only three or four objects within or just outside 100\,pc \citep{Harris18}.  The optical spectrum is spectacular and shows emission from higher Balmer lines as well as Ca\,{\sc ii} H and K \citep{Rossi2011}.  There is a detection of this source in {\em Galex} in the near-ultraviolet only, with $m_{\rm NUV}=20.3$\,AB\,mag.  Similar to the more detailed argument above for J0842, it is unlikely that emission from the dC star is responsible for near-ultraviolet flux detection, and in this case, stellar emission lines would need to be roughly 400 times brighter than the expected continuum.  Based solely on the absolute magnitude of this ultraviolet detection, and assuming it was due to a typical surface gravity white dwarf at the {\em Gaia} parallax distance to this target, it would imply $T_{\rm eff}\approx7500$\,K.  

\citet{Whitehouse18} reported several radial-velocity observations of this star that strongly indicate binarity at a significance just above $7\upsigma$.  Analysis of the previous and newer radial-velocity data with the \textsc{joker} yields only 70 surviving samples.  The subsequent MCMC posterior distribution for SBSS\,1310 is unimodal for all parameters, with the orbital period tightly constrained at 4.43\,d with a velocity semi-amplitude of $46.5\pm0.84$\,km\,s$^{-1}$.

The {\em TESS} data reveal a strong periodic signal corresponding to a photometric period of $5.26 \pm 0.01$\,d.  Unexpectedly, the photometric analysis on the individual, cleaned light-curves from each {\em TESS} sector reveals a clear change in the photometric period; Sectors 15, 16, and 22, individually, yield periods of $5.17 \pm 0.01$, $5.24 \pm 0.01$, and $5.30 \pm 0.01$\,d, respectively.  Also serendipitously, the amplitude of the photometric variation changes significantly between Sector 16 and 22, by around a factor of five.  These data are discussed in more detail in Section~\ref{subsec:drot}.

\subsection{CBS\,311}

This system is a previously known binary with a composite spectrum \citep{Liebert1994}, consisting of the dC star and a $T_{\rm eff}\approx31\,000\,$K DA ($\equiv$ strongest spectral lines are hydrogen) white dwarf.  The region of the sky occupied by this target lacks any {\em Galex} observations that could further constrain the luminosity of the companion.  CBS\,311 is one of the dC stars detected in X-rays by {\em Chandra} \citep{Green19}, but it should be noted again that the white dwarf cannot be the source of the high-energy emission.  This target was also imaged at high spatial resolution using the {\em Hubble Space Telescope}, where the binary remained unresolved to at least 25\,mas \citep{Farihi10}, corresponding to a projected separation on the sky of roughly 10\,AU at the 440\,pc {\em Gaia} DR2 parallax distance.  

The dC star in this binary is fainter than suggested by the $r$-band brightness due to some flux contribution from the white dwarf.  This diminishes the effective S/N of the absorption lines used for cross-correlation when measuring relative and absolute velocities.  Therefore, each spectrum of CBS\,311 was re-binned to a lower resolution by taking the mean of every 3 pixels.  The radial-velocity analysis identifies one prominent peak in the posterior distribution of the orbital parameters, at 0.19\,d with small peaks to either side, where the velocity semi-amplitude is constrained to be $29.2^{+2.0}_{-1.7}$\,km\,s$^{-1}$.  

The photometric analysis was conducted on the available ZTF data for this target.  A clear periodic signal corresponding to the strongest periodogram peak is located at $0.302\pm0.001$\,d, with a semi-amplitude near $\pm3$\,per cent.

\begin{figure*}
\includegraphics[width=\textwidth]{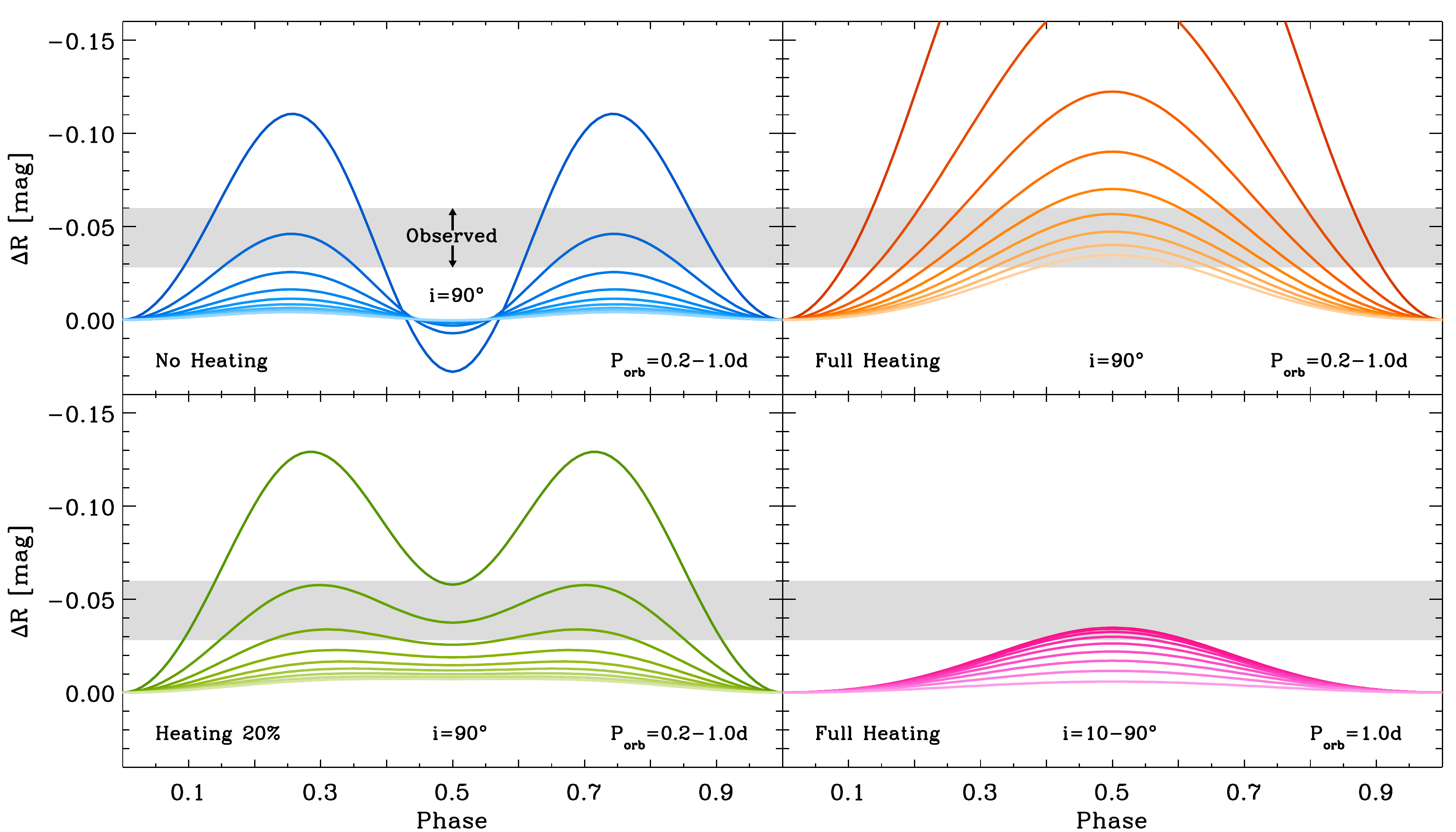}
\vskip -1 pt
\caption{Representative $R$-band light-curve simulations (detailed in Section~\ref{sec:binsimz}) for orbital periods representative of the shortest yet measured or suggested for dC stars.  The grey bar in each panel represents the range of peak-to-peak variations observed in the dC-star sample (0.028--0.060\,mag).  The top-left panel demonstrates the influence of tidal distortion alone, while the models represented at top right have identical parameters but include heating (irradiation).  The lower-left panel explores the effect of reduced heating (i.e., from cooler white-dwarf companions), and the bottom-right panel shows the effects of decreasing inclination.
\label{fig:Sims}}
\end{figure*}

\section{Discussion}

Spectroscopic and photometric periods are closely similar for all six systems for which both have been established, \lw{and} are indistinguishable for only two (Table~\ref{tab:Summary}).  For those two, orbital variability is, ostensibly, a plausible mechanism for the photometric changes.  For the remaining stars (at least), the working hypothesis is that the photometric variability must arise through rotational modulation, with the close similarities between orbital and rotational periods arising through tidal synchronisation.

In this section, binary light-curve simulations are presented, in order to provide context for the observed photometric variability.  Potential sources of the periodic flux changes, and their implications for the observed emission-line activity, are then explored.  Finally, the origin and evolution of these binaries are discussed in a wider context of carbon-enhanced stars.

\subsection{Binary light-curve simulations}
\label{sec:binsimz}

As an aid to interpreting the observed photometric variability, a range of orbital light-curves are simulated in order to gauge the possible effects of close-binary interaction.  The goal is to distinguish purely orbital photometric effects from the rotation of the dC star itself.  In particular, the aim is to characterise tidal-distortion and heating (irradiation, or day--night) effects arising from a white-dwarf companion, and thereby to characterise the parameter space where these effects may be significant.  

A standard Roche model is adopted, with simple `deep' heating, using the current version of a code originally described by \citet{Howarth82}.
These illustrative calculations are carried out for the Cousins $R$ band, as a rough match to the observed photometry, with the dC component modelled as a star with $T_{\rm dC}=4000$\,K, $M_{\rm dC}=0.44\,\Msun$, and $R_{\rm dC}=0.42\,\Rsun$ (if the lobe size allows; lobe-filling otherwise), which are approximately the values expected for a star near the bottom of the K-dwarf sequence for a metallicity $\text{[Fe/H]} = -1.0$ and an age of 1\,Gyr or older \citep{Dotter08}.  

The choice of metallicity has a strong influence on the inferred mass and radius for a fixed effective temperature, where going from solar metallicity to $\text{[Fe/H]} = -2.0$ changes the implied stellar mass from roughly 0.6 to $0.3\,\Msun$ at 4000\,K.  The metallicity of dC stars remains poorly determined, except in a single case of extreme metal poverty (G77-61 with [Fe/H] $= -4.0$; \citealt{Plez05}), yet the kinematics of dC stars clearly indicates that the population includes a large fraction of metal-poor members.  \lw{The adopted metallicity of [Fe/H] $=-1.0$ yields} $M_r =8.6$\,AB\,mag, and this compares favourably with the stars studied here, whose absolute magnitudes span $M_r =8.0$--9.2\,AB\,mag (see Table \ref{tab:Summary}).

 The adopted default white-dwarf parameters are $T_{\rm WD}=30\,000$\,K, $M_{\rm WD}=0.60\,\Msun$, $R_{\rm WD}=0.014\,\Rsun$, based on parameters for the white-dwarf companion to CBS\,311 \citep{Liebert1994}, one of only two known hot, bright companions to any dC star.
The implied mass--radius ratio is fairly typical of single white dwarfs, as well as those found binaries (although in the models the adopted radius is of consequence only for its role in determining the luminosity, $L/\Lsun \simeq 0.14$, which determines the magnitude of the irradiation).
A white dwarf of this nature gives rise to a significant heating (or day--night) effect on its companion.  In order to explore the effect of fainter white dwarfs, each simulation at a given orbital period was run for a range of heating efficiencies from 100\,per cent down to 20\,per cent, where the latter roughly corresponds to $T_{\rm WD}=20\,000$\,K for the same white-dwarf mass and radius.

The models incorporate a number of simplifications that are inconsequential for the purposes of this study.  The dC surface intensities are calculated using a constant, linear limb-darkening coefficient of 0.6 (roughly appropriate for the range of temperatures found), a gravity-darkening exponent of 0.08, and emergent fluxes (from \citealt{Howarth11}) interpolated at fixed $\log\,[g\,{\rm (cm\,s}^{-2})] = 5.0$.
While the white dwarf is bolometrically luminous (giving rise to heating of the dC-star atmosphere), it contributes no flux directly to the simulated light-curves.  Orbital periods are taken from the range $P_{\rm orb}=0.2$--1.0\,d -- the shortest measured or suggested periods known -- in steps 0.1\,d, with inclinations $i=10$--90$\degr$ in steps of $10\degr$.

Figure \ref{fig:Sims} shows a representative sample of the simulated light-curves, each panel holding some parameters fixed while varying others.  Notably for these parameters, the dC star fills its Roche lobe just below the simulation range at a period of 0.1\,d ($R \simeq 0.33\Rsun$).  At 0.2\,d, the tidal distortion of the star results in a photometric semi-amplitude near $\pm0.15$\,mag, and with the characteristic double-peaked, non-sinusoidal pattern.  Adding heating at the shortest simulated period results in a single peak and semi-amplitude just over $\pm0.30$\,mag.  At longer periods (i.e., wider separations), both of these orbital-motion-induced effects diminish rapidly; their full combined effect at 0.4\,d is less than half that at 0.2\,d.  For periods approaching and exceeding 1.0\,d, both tidal distortion and irradiation have a negligible impact on the emergent flux.

The magnitude of these calculated effects are taken here only as a representative guide, and they are accurate only to a degree.  For the tidal distortion, a classical Roche lobe is used for the equipotential surface, without the influence of asynchronous rotation.  The atmosphere of the irradiated star is only changed insofar as a local effective temperature modification, and without any calculation of the actual atmospheric structure.  Despite these potential shortcomings, the simulations should be sufficient to determine whether an observed photometric variation can plausibly be attributable to orbital motion, as opposed to stellar rotation.

\subsection{Source of photometric variability}

Here, the discussion pursues the question of whether the observed photometric variability in any given system can be attributed to orbital motion or to rotation of the dC star.

\subsubsection{Rotation versus orbit}

The seven dC stars listed in Table \ref{tab:Summary} exhibit photometric variability amplitudes falling between $\pm0.014$ and $\pm0.030$\,mag, with no apparent correlations with orbital period, or with (crudely estimated) inclination.  Each system is compared with the light-curve simulations (Fig.~\ref{fig:Sims}) for the appropriate orbital period (or range), and none are consistent with ellipsoidal variations induced by tidal distortion.  There are two elements to this inference; first, the amplitudes of the predicted changes are significantly smaller than those observed, and secondly, the dominant photometric period of ellipsoidal variations should be exactly half the orbital period.  While ellipsoidal variations would occur more readily in systems with shorter orbital periods, neither J1015 nor CBS\,311 (each with $P_{\rm orb}\approx0.2$\,d) exhibit such light-curve characteristics.  As a result of having well-constrained radial-velocity periods for six of the targets, it can be securely stated that none of their light-curves are consistent with an origin in tidal-distortions.

Next, the observed photometric changes are compared with those that might be induced due to irradiation from a relatively hot and luminous white-dwarf component -- the day--night, or heating, effect.  The predicted amplitudes can match or exceed the observed ranges, and so a more detailed examination is needed.  The simulations suggest CBS\,311 might experience a $\pm0.15$\,mag modulation in its light-curve due to irradiation, and the observed value is around fives times smaller than this, and thus is unlikely.  For J1015, the companion appears to be roughly three times less luminous (see below), and because the irradiated flux scales with luminosity, the heating effect will commensurately decrease.  In this case, the simulations indicate that a mildly double-peaked structure is expected, with a semi-amplitude near $\pm0.07$\,mag, and thus a possibility.  

There are two additional effects that would mitigate the magnitude of any orbital-induced photometric changes.  First, in the simulations, the white dwarf contributes no flux to the light-curve, and thus in cases where the white dwarf is visible in the optical spectrum, there will be a corresponding dilution factor from the contribution of its flux.  In CBS\,311, the two stars contribute nearly equal flux in the $r$ band \citep{Liebert1994}, and thus the predicted amplitude of the heating effect on the dC star will be diluted by a factor $\sim$2.  For J1015, the ultraviolet through optical spectra and photometry are decently fit by the addition of a $T_{\rm eff}=23\,500$\,K DA-type white dwarf, whose flux in the $r$ band is approximately 25\,per cent of the total, and the amplitude of any heating effect will be reduced accordingly.  Secondly, for an $i=0\degr$ (face-on) binary, there is no expected flux modulation from either tidal distortion or irradiation, as both binary components are in full view for the entire orbit.  The amplitude of both these orbital effects on the light-curve should increase to first order as $\sin{i}$ (Figure \ref{fig:Sims}), and thus in real observations, where present, these effects will typically be modestly decreased.

\begin{figure}
\includegraphics[width=\columnwidth]{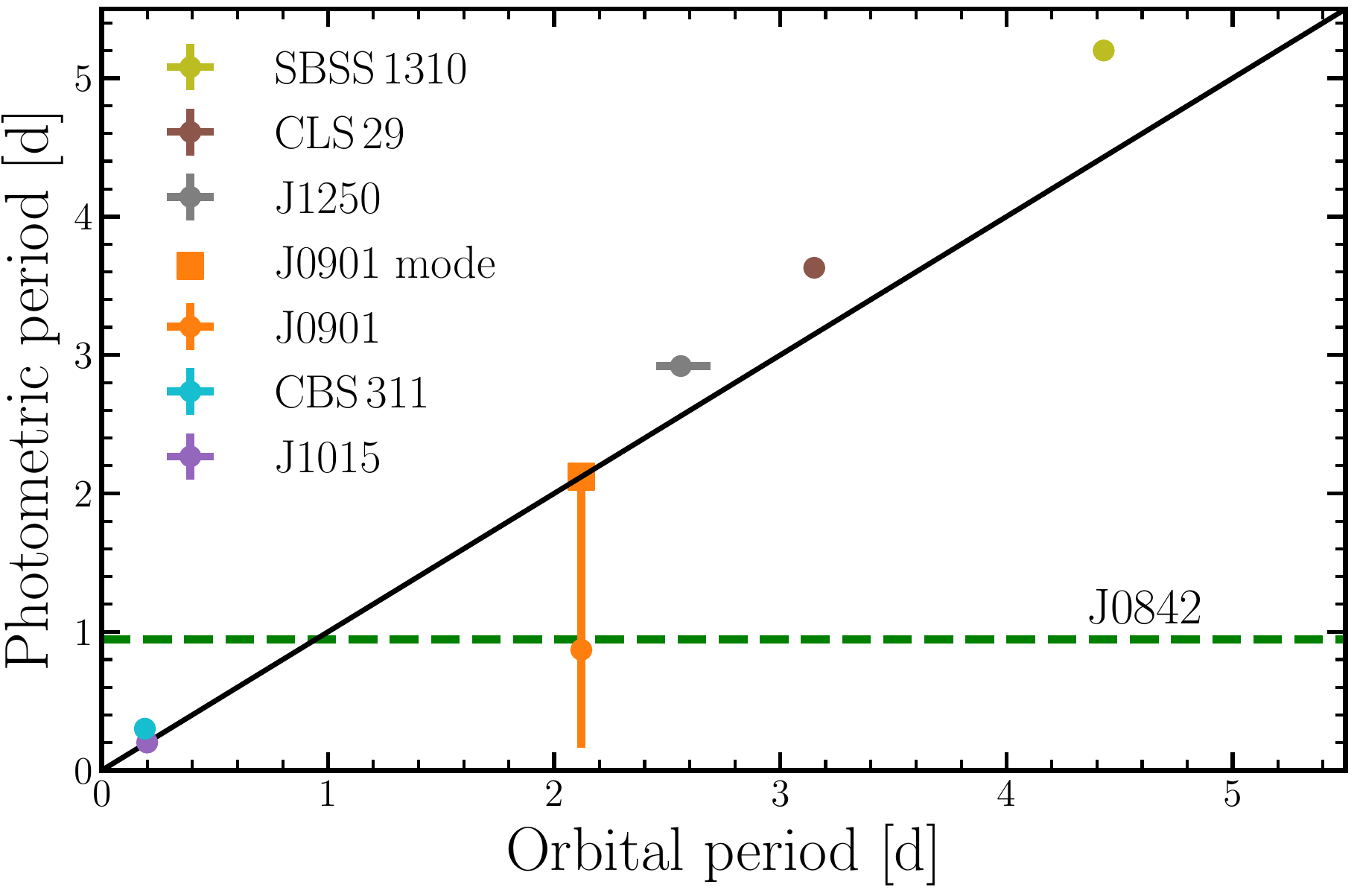}
\vskip -5 pt
\caption{The orbital period compared to the photometric period for the seven stars studied in this sample.  The error bars are shown as the 16\,per cent and 84\, per cent confidence limits.  Due to the highly non-Gaussian errors in the photometric period of J0901, the mode of the photometric period is also shown.  The solid black line represents the identity line between the orbital and photometric periods.  As there are no constraints on the orbital period of J0842, the photometric period is plotted as a green dashed line across all possible orbital periods. Each point is labelled with the corresponding name.}
\label{fig:Photvsrvp}
\end{figure}

Taking into account all of these factors as the data permit, the light-curves of both CBS\,311 and J1015 become inconsistent with irradiation for relatively low inclinations.  Specifically, for $i<42\degr$ the observed photometric variation of CBS\,311 is more consistent with rotation, and similarly for J1015 for $i<25\degr$.  For the stars J0842, J0901, CLS\,29, J1250, and SBSS\,1310, the only realistic source of photometric variability is rotation.  This is next considered as the primary or exclusive cause of the observed light-curve variations in the sample of dC stars is axial rotation.

\subsubsection{Tidal synchronisation}

Figure~\ref{fig:Photvsrvp} shows the relationship between the photometric and orbital periods for the six targets with well-constrained radial-velocity periods (including J1250).  Two systems have essentially identical photometric and radial-velocity periods, while the remaining four have photometric periods that are modestly longer than the radial-velocity periods (further undermining the possibility that the photometric variations are orbital).  
 
Those systems with matching photometric and radial-velocity periods -- J0901 and J1015 -- are neither the shortest period systems, nor those with the most luminous companions, as might be expected if tidal distortion or irradiation were responsible for the photometric variability.  The stars J1250, CLS\,29, SBSS\,1310, and CBS\,311 possess photometric periods on average around 15\,per cent longer than their orbital periods, which is strictly inconsistent with orbital motion.  

The collective data on these emission-line dC stars therefore broadly support a rotational origin for the photometric changes.  However, magnetic braking causes spin-down with age for isolated convective stars.  Assuming a crude, and probably conservative, age estimate for dC stars of a few to several Gyr \citep{Farihi18}, the empirical age--rotation relationship for single M dwarfs predicts spin periods on the order of 20--100\,d \citep{Engle_2018} -- significantly longer than any of the photometric periods measured in this study, implying an additional source of spin angular momentum.  

Tidally induced rotational synchronisation is an appealing mechanism for the implied spin-up and relatively rapid rotation, particularly since the radial-velocity data are consistent with circular orbits (and rotational synchronisation is expected to take place on shorter timescales than orbital circularisation).
Theoretical timescales for synchronisation depend mainly on the ratio of the radius of the star experiencing tidal forces (here the dC star) to the orbital separation between the binary components.  Adopting canonical values for M-dwarf radii, along with the orbital separations determined from this study, it is found that synchronisation timescales are a few Myr up to no more than a Gyr \citep{Zahn89,Zahn}.  

Assuming that all of the binary companions are now white dwarfs, their cooling timescales can be estimated and provide a window over which tidal synchronisation could have occurred.  Attributing all {\em Galex} ultraviolet flux in these systems to white dwarfs -- and upper limits for non-detections -- at the {\em Gaia} distances to the dC stars, and assuming a fixed white-dwarf surface gravity of $\log\,[g\,{\rm (cm\,s}^{-2})] = 8.0$, cooling-age estimates and lower limits can be derived from models \citep{B_dard_2020}.  The resulting cooling ages range from approximately 10\,Myr (CBS\,311) to over 1\,Gyr, but in all cases are broadly consistent with the predicted synchronisation timescale.

\subsubsection{Differential rotation}
\label{subsec:drot}

Of course, for fully synchronised rotation the photometric signal should exactly match the radial-velocity period (as assumed in the analysis of J1250 by \citealt{Margon18}).  However, there are modest offsets between the orbital and photometric periods for much of the sample, with the photometric period never shorter than the radial-velocity period.  To reconcile this result with fully synchronised \textit{equatorial} rotation, the dC star must rotate differentially, with starspots 
(or other surface-brightness inhomogeneities) at relatively high latitudes; for example, solar differential rotation implies a 15\,per cent longer period at latitudes of 20--35\degr than at the equator.

\begin{figure*}
\includegraphics[width=\textwidth]{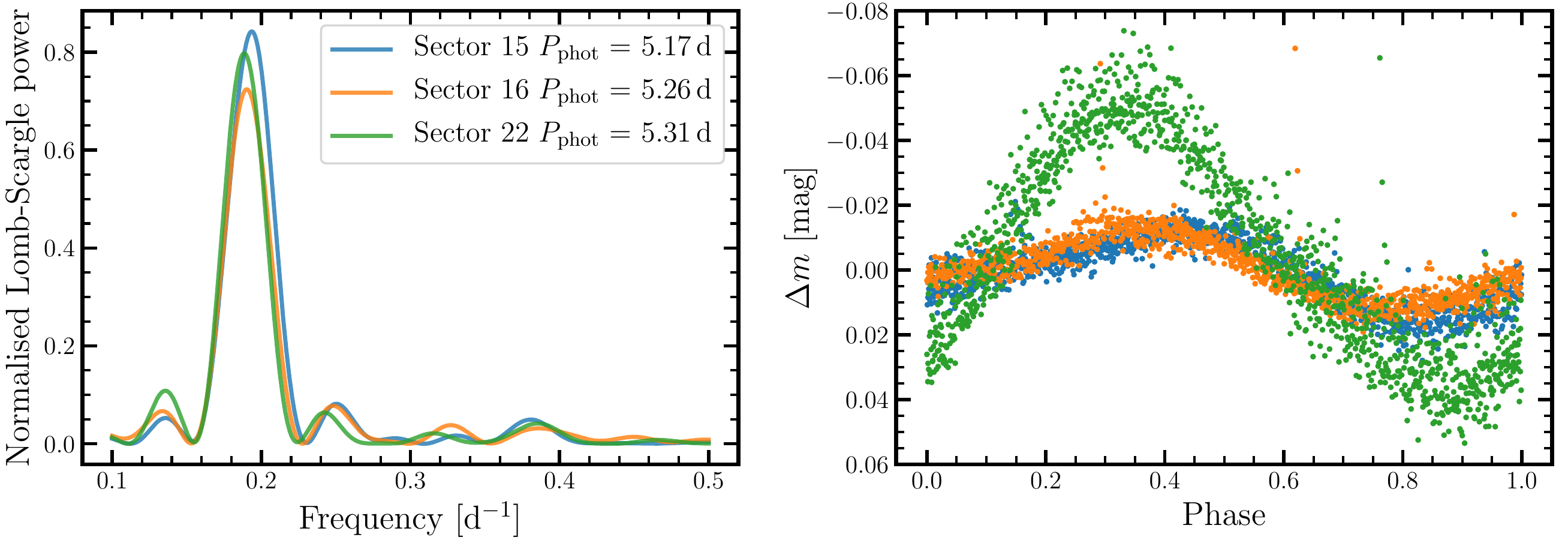}
\vskip -5 pt
\caption{\textit{Left:} The normalised LS periodograms computed from the individual \textit{TESS} sectors for SBSS\,1310. \textit{Right:} The light-curves of each sector, phase-folded to the 5.26\,d period determined from the analysis of all \textit{TESS} sectors combined, and using the time of the first observation as the arbitrary `phase zero'.  There is a clear increase in amplitude for Sector 22 relative to the first two observed sectors, completed 139 days prior.  Changes in the light-curve shape and period result in the offsets in phase of maximum brightness.}
\label{fig:SBSS}
\end{figure*}

The plausibility of differential rotation is best illustrated by SBSS\,1310. As noted in Section~\ref{subsec:sbssp}, the photometric period of this star shows modest but significant changes between \textit{TESS} Sectors 15, 16, and 22 (5.17, 5.24, and 5.30\,d, respectively, with $\upsigma=0.01$\,d uncertainties).  The individual sector periodograms are presented in Figure~\ref{fig:SBSS}, together with the sector light-curves phase-folded to the mean 5.26\,d period.  Evidently, the light-curve changes not only in period, but also in shape and amplitude.  The simplest interpretation of these characteristics is changes in the geometry of starspots (resulting in the variable light-curve shape and amplitude) and in their latitude, accompanied by differential rotation (resulting in the variable period).   Differential rotation also accommodates the difference between the radial-velocity and mean photometric periods (4.4\,d vs.\, 5.26\,d), suggesting a star-spot latitude of $\gtrsim$20\degr (assuming solar-type differential rotation and equatorial synchronisation).  

If SBSS\,1310 is representative of its spectral sub-class, then the targets appear to be differentially rotating, tidally synchronised stars.

\subsection{Source of stellar activity}

There are two basic possibilities for the source of stellar activity in dC stars, as manifested by H$\upalpha$ and similar emission features; an intrinsic source via relatively rapid stellar rotation, and an extrinsic source via irradiation.  

In order to heat a low-mass star to sufficiently high local surface temperatures that emission lines are excited, a white-dwarf companion must be close to the irradiated star and fairly luminous.  Based on a comprehensive study of stellar activity as traced by H$\upalpha$ emission in M dwarf companions to white dwarfs \citep{Rebassa13}, it was found that activity is nearly always intrinsic for both partly and fully convective stars.  Furthermore, while the influence of magnetic braking is seen for partly convective stars in wide binaries, those M dwarfs in short-period orbits resulting from a common envelope all remain active.  This fact alone argues that irradiation is not the cause of stellar activity in low-mass stars closely orbiting white dwarfs, and the study furthermore demonstrates that only white dwarfs with an irradiating flux that is at least $3\times$ higher than the intrinsic M dwarf flux can result in emission \citep{Rebassa13}.  None of the seven short-period dC stars in this study approaches this level of irradiation; e.g.\ the 30\,000\,K white dwarf in CBS\,311 is only $1.5\times$ more luminous than its companion, and thus falls drastically short of this criterion at the surface of the dC star.  Therefore, irradiation cannot be responsible for the observed activity in dC stars.

The remaining possibility requires relatively rapid rotation to generate a solar-type dynamo \citep{Parker55}, and an active chromosphere \citep{chromo}.  The fact that dC stars with emission lines are relatively rapid rotators is clearly evidenced by X-ray detections in six out of six observed cases \citep{Green19}, including four of the stars analysed here.  There are three potential angular momentum sources for the rapid rotation observed in dC stars: intrinsic spin of formation, spin-up owing to mass transfer, and tidal synchronisation in a binary.  

Three lines of evidence discount the idea that dC stars are relatively young and hence, still rotating rapidly due to their youth.  One is their overall kinematical properties that suggest a mixture of old disc and halo stars \citep{Farihi18}, second is the indication from population synthesis that metal-poor stars are more readily polluted to C/O $>1$ than solar abundance stars \citep{Kool95}, and third is through estimating the cooling ages of their white-dwarf companions.  The actual masses and temperatures of dC stars are certainly not well constrained, and not established yet at any level of confidence, but based on their positions on the H-R Diagram they should be partly convective stars with parameters comparable to late-type dwarfs, and because of inevitable magnetic braking \citep{Skum,McQuil}, they require an extrinsic origin to their current spins, assuming their total ages are at least a few Gyr.

Thus, only mass transfer or tidal synchronisation remain as possibilities for the observed, spin-induced stellar activity.  And it should be noted that both can occur, that is they are not mutually exclusive (however, it is noteworthy that the latter can erase any evidence of the former, e.g.\ in the case the dC was once spinning faster than the orbital period).  If some fraction of the emission-line dC stars are the product of spin-up through mass transfer, and not the result of tidal synchronisation in short-period binaries, the observed rotation rate would be an indication of how much mass the system has accreted.  Therefore, measurements of the rotation rates in any systems that are not short-period binaries \lw{(such as for the Ba giant HD\,165141; \citealt{Theuns96})} would provide constraints on the amount of mass, and angular momentum accreted, and thus inform mass transfer models \citep{Izzard10,Mat17}.  

\subsection{Period distribution}

The overall results indicate clearly that H$\upalpha$ emission in dC stars is linked to short-period binarity.  In fact, of the ten dC stars with constrained orbital parameters, there are two distinct families: those with H$\upalpha$ emission all have orbital periods of a few days or less, those without detectable H$\upalpha$ emission have orbital periods of several months or longer \citep{Dearborn86,Harris18,Margon18,Roulston2021}.

The data at hand suggest that the orbital period distribution of the dC star population may be bimodal, and thus consistent in some key aspects with the (vastly) more commonly occurring M-type dwarf companions to white dwarfs.  In general, such bimodal populations are consistent with the post-main sequence removal of intermediate-period orbits, via a combination of common-envelope evolution for shorter periods, and the reduction of gravitational binding energy due to mass loss for longer periods.  However, there is a notable difference in the dC star binary population, and that is the orbital periods on the order of years, which are absent in the carbon-normal, M-dwarf companion population orbiting white dwarfs \citep{Farihi10,Nebot2011,Ashley2019}.  It is tempting to speculate that the dC companion orbits may differ from dM companion orbits owing to mass transfer, but that is beyond the scope of the current study.  Nevertheless it is clear that the short-period dC binaries have experienced a common envelope \citep{Izzard12,Ivanova13}.  If the emission-line dC stars have evolved through a common-envelope phase, and assuming their carbon enhancement is extrinsic, then this suggests relatively stable mass transfer {\em prior to the common envelope}.  

Extrinsically carbon-enhanced binaries possessing short-orbital periods are rare.  Amongst the Ba, CH, and CEMP-s populations, which typically possess orbital periods of many months or \lw{years \citep{Cemps,CH,Jorissen19},} only two stars are known to have orbital periods shorter than ten days.  One of these is the central binary in the Necklace nebula, with a 1.2\,d orbital period, and whose exact link to other carbon-enhanced populations remains to be determined \citep{Miszalski13}.  The other is HE\,0024-2523, a 0.9\,\Msun main-sequence star with a 3.4\,d orbital period measured through radial-velocity monitoring \citep{Luc03}, and a CH spectral class.  Similar to the short-period binaries found here, HE\,0024-2523 is suspected of receiving mass transfer from an evolved companion before evolving through a common envelope.  However, there is no evidence of short-period photometric variability or rotation in HE\,0024-2523, despite decent quality {\em TESS} data, and the star lacks emission lines.

It is tempting to speculate why, on the one hand, Ba, CH, and CEMP-s stars appear to result from relatively wide and essentially detached binary evolution; and on the other hand, why the vast majority of post-common envelope companions to white dwarfs appear to be carbon-normal M dwarfs.  Unfortunately, there are currently no robust constraints on the fundamental stellar parameters of dC stars (i.e.\ mass, effective temperature, metallicity), and a distinct lack of appropriate atmospheric models.  It is therefore beyond the scope of the current study to address the different pathways in stellar and binary evolution that might give rise to the observed orbital period differences in these populations.  However, it is hoped that the empirical results presented here will encourage novel evolutionary modeling.  Because dC stars should far outnumber their evolved counterparts, those with short orbital periods provide valuable constraints for models of extrinsic carbon-enrichment.

\section{Conclusions}

This study analyses time-series radial-velocity and photometric data for seven dC stars with H$\upalpha$ emission lines.  Six of the dC stars analysed here are found to possess orbital periods in the range 0.2--4.4\,d, along with photometric periods that are either identical (in the case of J0901 and J1015), or within a few tens of per cent of the orbital period.  The seventh dC star, J0842, lacks significant radial-velocity variations but possesses a 0.945\,d photometric period, and if these periods are similar it would be difficult to detect through ground-based radial-velocity measurements.  

Binary light-curve simulations suggest that the origin of the photometric variations observed in these emission-line dC stars is consistent with rotation.  None of the light-curves show any evidence for a day- and night-side effect or tidal distortion, and while irradiation is physically plausible in two of the seven systems, it is unlikely based on binary inclination and companion brightness estimates.  Thus, rotation is the most likely cause for the observed photometric variations.  The mostly sinusoidal behavior of the (best) observed light curves is not necessarily expected for a random assortment of starspot latitudes and stellar spin inclinations, but both the light-curve data quality and the number of short-spin period dC stars are insufficient to determine further constraints on the geometry and number of starspots.

The expectations of magnetic braking are such that the dC stars should be spinning an order of magnitude more slowly than observed, and their ages are consistent with sufficient time to have become tidally synchronised to their visible or unseen (and presumed to be) white-dwarf companions.  Tidal synchronisation provides the most consistent picture for these short-period dC star binaries, and provides the additional angular momentum needed for spin-up, which in turn powers the emission lines through a solar-type dynamo.  However, mass transfer cannot be ruled out, but in the sample studied here, mass transfer is only necessary for the extrinsic source of pollution, and not for spin-up.  Lastly, at least a few of the dC stars appear to be differentially rotating, as evidenced by photometric periods that are similar to, but somewhat longer than their orbital periods, as well as at least one system with changing photometric period and variability amplitude.

This study more than doubles the number of dC stars with constrained orbital periods, and increases those known to be in short-period orbits from one to six.  It appears the dC binary period distribution may be bimodal, with a post-common-envelope population as studied here, and a wider population with periods on the order of months to years.  Interestingly, the other well-studied, carbon-enhanced stellar populations lack a short-period, post-common-envelope population, and it is likely the dC stars will provide key insight into mass transfer models on the basis of these, and future orbital period determinations.

\section*{Acknowledgements}

LW would like to thank P.~Jethwa and S.~Lilleengen for useful discussion relating to the analysis in this paper, and the reviewer, Alain Jorissen, for a positive and insightful report. LW also acknowledges support from the ESO studentship programme, and STFC studentship.  The radial-velocity data obtained in this paper was taken using the William Herschel Telescope that is operated on the island of La Palma by the Isaac Newton Group of Telescopes in the Spanish Observatorio del Roque de los Muchachos of the Instituto de Astrof\'isica de Canarias.  This paper includes data collected by the {\em TESS} and \textit{K2} missions, with funding for these missions provided by the NASA Explorer Program and NASA Science Mission directorate.

\textit{Software:} \textsc{Astropy} \citep{Astropy}, \textsc{Corner} \citep{corner}, \textsc{iraf} \citep{Iraf}, \textsc{Joker} \citep{Joker}, \textsc{Matplotlib} \citep{Matplotlib}, \textsc{Numpy} \citep{Numpy}, \textsc{p4j} \citep{P4J}, \textsc{Pandas} \citep{pandas}, \textsc{Scipy} \citep{Scipy}

\section*{Data availability statement}

\textit{TESS} and \textit{K2} data used in this article are available via the Mikulski Archive for Space Telescopes portal, maintained and funded by NASA. ZTF data are available to download in the NASA/IPAC infrared science archive. Finally, ISIS spectroscopy are available at the Isaac Newton Group Archive which is maintained as part of the CASU Astronomical Data Centre at the Institute of Astronomy, Cambridge.




\bibliographystyle{mnras}
\bibliography{refs} 

\begin{thebibliography}{}
\makeatletter
\relax
\def\mn@urlcharsother{\let\do\@makeother \do\$\do\&\do\#\do\^\do\_\do\%\do\~}
\def\mn@doi{\begingroup\mn@urlcharsother \@ifnextchar [ {\mn@doi@}
  {\mn@doi@[]}}
\def\mn@doi@[#1]#2{\def\@tempa{#1}\ifx\@tempa\@empty \href
  {http://dx.doi.org/#2} {doi:#2}\else \href {http://dx.doi.org/#2} {#1}\fi
  \endgroup}
\def\mn@eprint#1#2{\mn@eprint@#1:#2::\@nil}
\def\mn@eprint@arXiv#1{\href {http://arxiv.org/abs/#1} {{\tt arXiv:#1}}}
\def\mn@eprint@dblp#1{\href {http://dblp.uni-trier.de/rec/bibtex/#1.xml}
  {dblp:#1}}
\def\mn@eprint@#1:#2:#3:#4\@nil{\def\@tempa {#1}\def\@tempb {#2}\def\@tempc
  {#3}\ifx \@tempc \@empty \let \@tempc \@tempb \let \@tempb \@tempa \fi \ifx
  \@tempb \@empty \def\@tempb {arXiv}\fi \@ifundefined
  {mn@eprint@\@tempb}{\@tempb:\@tempc}{\expandafter \expandafter \csname
  mn@eprint@\@tempb\endcsname \expandafter{\@tempc}}}

\bibitem[\protect\citeauthoryear{{Abate}, {Pols}, {Izzard}, {Mohamed}  \& {de
  Mink}}{{Abate} et~al.}{2013}]{Abate13}
{Abate} C.,  {Pols} O.~R.,  {Izzard} R.~G.,  {Mohamed} S.~S.,   {de Mink}
  S.~E.,  2013, \mn@doi [\aap] {10.1051/0004-6361/201220007}, \href
  {https://ui.adsabs.harvard.edu/abs/2013A&A...552A..26A} {552, A26}

\bibitem[\protect\citeauthoryear{{Abbott} et~al.,}{{Abbott}
  et~al.}{2016}]{Grav_wav}
{Abbott} B.~P.,  et~al., 2016, \mn@doi [\prl] {10.1103/PhysRevLett.116.061102},
  \href {https://ui.adsabs.harvard.edu/abs/2016PhRvL.116f1102A} {116, 061102}

\bibitem[\protect\citeauthoryear{{Ashley}, {Farihi}, {Marsh}, {Wilson}  \&
  {G{\"a}nsicke}}{{Ashley} et~al.}{2019}]{Ashley2019}
{Ashley} R.~P.,  {Farihi} J.,  {Marsh} T.~R.,  {Wilson} D.~J.,   {G{\"a}nsicke}
  B.~T.,  2019, \mn@doi [\mnras] {10.1093/mnras/stz298}, \href
  {https://ui.adsabs.harvard.edu/abs/2019MNRAS.484.5362A} {484, 5362}

\bibitem[\protect\citeauthoryear{{Astropy Collaboration} et~al.,}{{Astropy
  Collaboration} et~al.}{2018}]{Astropy}
{Astropy Collaboration} et~al., 2018, \mn@doi [\aj] {10.3847/1538-3881/aabc4f},
  \href {https://ui.adsabs.harvard.edu/abs/2018AJ....156..123A} {156, 123}

\bibitem[\protect\citeauthoryear{B{\'{e}}dard, Bergeron, Brassard  \&
  Fontaine}{B{\'{e}}dard et~al.}{2020}]{B_dard_2020}
B{\'{e}}dard A.,  Bergeron P.,  Brassard P.,   Fontaine G.,  2020, \mn@doi [The
  Astrophysical Journal] {10.3847/1538-4357/abafbe}, 901, 93

\bibitem[\protect\citeauthoryear{{Bellm}}{{Bellm}}{2014}]{Bellm2014}
{Bellm} E.,  2014, in {Wozniak} P.~R.,  {Graham} M.~J.,  {Mahabal} A.~A.,
  {Seaman} R.,  eds, The Third Hot-wiring the Transient Universe Workshop. pp
  27--33 (\mn@eprint {arXiv} {1410.8185})

\bibitem[\protect\citeauthoryear{{Boro Saikia} et~al.,}{{Boro Saikia}
  et~al.}{2018}]{chromo}
{Boro Saikia} S.,  et~al., 2018, \mn@doi [\aap] {10.1051/0004-6361/201629518},
  \href {https://ui.adsabs.harvard.edu/abs/2018A&A...616A.108B} {616, A108}

\bibitem[\protect\citeauthoryear{{Castelli} \& {Kurucz}}{{Castelli} \&
  {Kurucz}}{2003}]{Kurucz03}
{Castelli} F.,  {Kurucz} R.~L.,  2003, in {Piskunov} N.,  {Weiss} W.~W.,
  {Gray} D.~F.,  eds,  IAU Symposium Vol. 210, Modelling of Stellar
  Atmospheres. p.~A20 (\mn@eprint {arXiv} {astro-ph/0405087})

\bibitem[\protect\citeauthoryear{{Dahn}, {Liebert}, {Kron}, {Spinrad}  \&
  {Hintzen}}{{Dahn} et~al.}{1977}]{Dahn77}
{Dahn} C.~C.,  {Liebert} J.,  {Kron} R.~G.,  {Spinrad} H.,   {Hintzen} P.~M.,
  1977, \mn@doi [\apj] {10.1086/155518}, \href
  {https://ui.adsabs.harvard.edu/abs/1977ApJ...216..757D} {216, 757}

\bibitem[\protect\citeauthoryear{{\DE{De Kool}{De}{de}}~Kool \&
  {Green}}{{\DE{De Kool}{De}{de}}~Kool \& {Green}}{1995}]{Kool95}
{\DE{De Kool}{De}{de}}~Kool M.,  {Green} P.~J.,  1995, \mn@doi [\apj]
  {10.1086/176051}, \href
  {https://ui.adsabs.harvard.edu/abs/1995ApJ...449..236D} {449, 236}

\bibitem[\protect\citeauthoryear{{Dearborn}, {Liebert}, {Aaronson}, {Dahn},
  {Harrington}, {Mould}  \& {Greenstein}}{{Dearborn} et~al.}{1986}]{Dearborn86}
{Dearborn} D.~S.~P.,  {Liebert} J.,  {Aaronson} M.,  {Dahn} C.~C.,
  {Harrington} R.,  {Mould} J.,   {Greenstein} J.~L.,  1986, \mn@doi [\apj]
  {10.1086/163805}, \href
  {https://ui.adsabs.harvard.edu/abs/1986ApJ...300..314D} {300, 314}

\bibitem[\protect\citeauthoryear{{Dotter}, {Chaboyer}, {Jevremovi{\'c}},
  {Kostov}, {Baron}  \& {Ferguson}}{{Dotter} et~al.}{2008}]{Dotter08}
{Dotter} A.,  {Chaboyer} B.,  {Jevremovi{\'c}} D.,  {Kostov} V.,  {Baron} E.,
  {Ferguson} J.~W.,  2008, \mn@doi [\apjs] {10.1086/589654}, \href
  {https://ui.adsabs.harvard.edu/abs/2008ApJS..178...89D} {178, 89}

\bibitem[\protect\citeauthoryear{Engle \& Guinan}{Engle \&
  Guinan}{2018}]{Engle_2018}
Engle S.~G.,  Guinan E.~F.,  2018, \mn@doi [Research Notes of the {AAS}]
  {10.3847/2515-5172/aab1f8}, 2, 34

\bibitem[\protect\citeauthoryear{{Farihi}, {Hoard}  \& {Wachter}}{{Farihi}
  et~al.}{2010}]{Farihi10}
{Farihi} J.,  {Hoard} D.~W.,   {Wachter} S.,  2010, \mn@doi [\apjs]
  {10.1088/0067-0049/190/2/275}, \href
  {https://ui.adsabs.harvard.edu/abs/2010ApJS..190..275F} {190, 275}

\bibitem[\protect\citeauthoryear{{Farihi}, {Arendt}, {Machado}  \&
  {Whitehouse}}{{Farihi} et~al.}{2018}]{Farihi18}
{Farihi} J.,  {Arendt} A.~R.,  {Machado} H.~S.,   {Whitehouse} L.~J.,  2018,
  \mn@doi [\mnras] {10.1093/mnras/sty890}, \href
  {https://ui.adsabs.harvard.edu/abs/2018MNRAS.477.3801F} {477, 3801}

\bibitem[\protect\citeauthoryear{{Fontaine}, {Brassard}  \&
  {Bergeron}}{{Fontaine} et~al.}{2001}]{Fontaine01}
{Fontaine} G.,  {Brassard} P.,   {Bergeron} P.,  2001, \mn@doi [\pasp]
  {10.1086/319535}, \href
  {https://ui.adsabs.harvard.edu/abs/2001PASP..113..409F} {113, 409}

\bibitem[\protect\citeauthoryear{Foreman-Mackey}{Foreman-Mackey}{2016}]{corner}
Foreman-Mackey D.,  2016, \mn@doi [The Journal of Open Source Software]
  {10.21105/joss.00024}, 1, 24

\bibitem[\protect\citeauthoryear{{Gaia Collaboration}, {Brown}, {Vallenari},
  {Prusti}, {de Bruijne}, {Babusiaux}  \& {Biermann}}{{Gaia Collaboration}
  et~al.}{2020}]{eDR3}
{Gaia Collaboration} {Brown} A.~G.~A.,  {Vallenari} A.,  {Prusti} T.,  {de
  Bruijne} J.~H.~J.,  {Babusiaux} C.,   {Biermann} M.,  2020, arXiv e-prints,
  \href {https://ui.adsabs.harvard.edu/abs/2020arXiv201201533G} {p.
  arXiv:2012.01533}

\bibitem[\protect\citeauthoryear{{Green}}{{Green}}{2013}]{Green2013}
{Green} P.,  2013, \mn@doi [\apj] {10.1088/0004-637X/765/1/12}, \href
  {https://ui.adsabs.harvard.edu/abs/2013ApJ...765...12G} {765, 12}

\bibitem[\protect\citeauthoryear{{Green} et~al.,}{{Green}
  et~al.}{2019}]{Green19}
{Green} P.~J.,  et~al., 2019, \mn@doi [\apj] {10.3847/1538-4357/ab2bf4}, \href
  {https://ui.adsabs.harvard.edu/abs/2019ApJ...881...49G} {881, 49}

\bibitem[\protect\citeauthoryear{{Gustafsson}, {Edvardsson}, {Eriksson},
  {J{\o}rgensen}, {Nordlund}  \& {Plez}}{{Gustafsson} et~al.}{2008}]{Marcs08}
{Gustafsson} B.,  {Edvardsson} B.,  {Eriksson} K.,  {J{\o}rgensen} U.~G.,
  {Nordlund} {\r{A}}.,   {Plez} B.,  2008, \mn@doi [\aap]
  {10.1051/0004-6361:200809724}, \href
  {https://ui.adsabs.harvard.edu/abs/2008A&A...486..951G} {486, 951}

\bibitem[\protect\citeauthoryear{{Harris} et~al.,}{{Harris}
  et~al.}{1998}]{Harris98}
{Harris} H.~C.,  et~al., 1998, \mn@doi [\apj] {10.1086/305908}, \href
  {https://ui.adsabs.harvard.edu/abs/1998ApJ...502..437H} {502, 437}

\bibitem[\protect\citeauthoryear{{Harris} et~al.,}{{Harris}
  et~al.}{2018}]{Harris18}
{Harris} H.~C.,  et~al., 2018, \mn@doi [\aj] {10.3847/1538-3881/aac100}, \href
  {https://ui.adsabs.harvard.edu/abs/2018AJ....155..252H} {155, 252}

\bibitem[\protect\citeauthoryear{Harris et~al.,}{Harris et~al.}{2020}]{Numpy}
Harris C.~R.,  et~al., 2020, \mn@doi [Nature] {10.1038/s41586-020-2649-2}, 585,
  357

\bibitem[\protect\citeauthoryear{{Heber}, {Bade}, {Jordan}  \& {Voges}}{{Heber}
  et~al.}{1993}]{Heber93}
{Heber} U.,  {Bade} N.,  {Jordan} S.,   {Voges} W.,  1993, \aap, \href
  {https://ui.adsabs.harvard.edu/abs/1993A&A...267L..31H} {267, L31}

\bibitem[\protect\citeauthoryear{{Hoffman} \& {Gelman}}{{Hoffman} \&
  {Gelman}}{2011}]{NUTS}
{Hoffman} M.~D.,  {Gelman} A.,  2011, arXiv e-prints, \href
  {https://ui.adsabs.harvard.edu/abs/2011arXiv1111.4246H} {p. arXiv:1111.4246}

\bibitem[\protect\citeauthoryear{{Howarth}}{{Howarth}}{1982}]{Howarth82}
{Howarth} I.~D.,  1982, \mn@doi [\mnras] {10.1093/mnras/198.2.289}, \href
  {https://ui.adsabs.harvard.edu/abs/1982MNRAS.198..289H} {198, 289}

\bibitem[\protect\citeauthoryear{{Howarth}}{{Howarth}}{2011}]{Howarth11}
{Howarth} I.~D.,  2011, \mn@doi [\mnras] {10.1111/j.1365-2966.2011.18122.x},
  \href {https://ui.adsabs.harvard.edu/abs/2011MNRAS.413.1515H} {413, 1515}

\bibitem[\protect\citeauthoryear{{Howell} et~al.,}{{Howell}
  et~al.}{2014}]{Howell14}
{Howell} S.~B.,  et~al., 2014, \mn@doi [\pasp] {10.1086/676406}, \href
  {https://ui.adsabs.harvard.edu/abs/2014PASP..126..398H} {126, 398}

\bibitem[\protect\citeauthoryear{{Huijse}, {Est{\'e}vez}, {F{\"o}rster},
  {Daniel}, {Connolly}, {Protopapas}, {Carrasco}  \& {Pr{\'\i}ncipe}}{{Huijse}
  et~al.}{2018}]{P4J}
{Huijse} P.,  {Est{\'e}vez} P.~A.,  {F{\"o}rster} F.,  {Daniel} S.~F.,
  {Connolly} A.~J.,  {Protopapas} P.,  {Carrasco} R.,   {Pr{\'\i}ncipe} J.~C.,
  2018, \mn@doi [\apjs] {10.3847/1538-4365/aab77c}, \href
  {https://ui.adsabs.harvard.edu/abs/2018ApJS..236...12H} {236, 12}

\bibitem[\protect\citeauthoryear{Hunter}{Hunter}{2007}]{Matplotlib}
Hunter J.~D.,  2007, \mn@doi [Computing in Science \& Engineering]
  {10.1109/MCSE.2007.55}, 9, 90

\bibitem[\protect\citeauthoryear{{Hurley}, {Tout}  \& {Pols}}{{Hurley}
  et~al.}{2002}]{Hurley02}
{Hurley} J.~R.,  {Tout} C.~A.,   {Pols} O.~R.,  2002, \mn@doi [\mnras]
  {10.1046/j.1365-8711.2002.05038.x}, \href
  {https://ui.adsabs.harvard.edu/abs/2002MNRAS.329..897H} {329, 897}

\bibitem[\protect\citeauthoryear{{Iben} \& {Renzini}}{{Iben} \&
  {Renzini}}{1983}]{Iben}
{Iben} I. J.,  {Renzini} A.,  1983, \mn@doi [\araa]
  {10.1146/annurev.aa.21.090183.001415}, \href
  {https://ui.adsabs.harvard.edu/abs/1983ARA&A..21..271I} {21, 271}

\bibitem[\protect\citeauthoryear{{Ivanova} et~al.,}{{Ivanova}
  et~al.}{2013}]{Ivanova13}
{Ivanova} N.,  et~al., 2013, \mn@doi [\aapr] {10.1007/s00159-013-0059-2}, \href
  {https://ui.adsabs.harvard.edu/abs/2013A&ARv..21...59I} {21, 59}

\bibitem[\protect\citeauthoryear{{Izzard}, {Dermine}  \& {Church}}{{Izzard}
  et~al.}{2010}]{Izzard10}
{Izzard} R.~G.,  {Dermine} T.,   {Church} R.~P.,  2010, \mn@doi [\aap]
  {10.1051/0004-6361/201015254}, \href
  {https://ui.adsabs.harvard.edu/abs/2010A&A...523A..10I} {523, A10}

\bibitem[\protect\citeauthoryear{{Izzard}, {Hall}, {Tauris}  \&
  {Tout}}{{Izzard} et~al.}{2012}]{Izzard12}
{Izzard} R.~G.,  {Hall} P.~D.,  {Tauris} T.~M.,   {Tout} C.~A.,  2012, in IAU
  Symposium. pp 95--102, \mn@doi{10.1017/S1743921312010769}

\bibitem[\protect\citeauthoryear{{Jorissen} et~al.,}{{Jorissen}
  et~al.}{2016}]{CH}
{Jorissen} A.,  et~al., 2016, \mn@doi [\aap] {10.1051/0004-6361/201526992},
  \href {https://ui.adsabs.harvard.edu/abs/2016A&A...586A.158J} {586, A158}

\bibitem[\protect\citeauthoryear{{Jorissen}, {Boffin}, {Karinkuzhi}, {Van Eck},
  {Escorza}, {Shetye}  \& {Van Winckel}}{{Jorissen} et~al.}{2019}]{Jorissen19}
{Jorissen} A.,  {Boffin} H.~M.~J.,  {Karinkuzhi} D.,  {Van Eck} S.,  {Escorza}
  A.,  {Shetye} S.,   {Van Winckel} H.,  2019, \mn@doi [\aap]
  {10.1051/0004-6361/201834630}, \href
  {https://ui.adsabs.harvard.edu/abs/2019A&A...626A.127J} {626, A127}

\bibitem[\protect\citeauthoryear{{Koester}}{{Koester}}{2010}]{Koester10}
{Koester} D.,  2010, \memsai, \href
  {https://ui.adsabs.harvard.edu/abs/2010MmSAI..81..921K} {81, 921}

\bibitem[\protect\citeauthoryear{{Kulkarni}}{{Kulkarni}}{2013}]{Kulkarni2013}
{Kulkarni} S.~R.,  2013, The Astronomer's Telegram, \href
  {https://ui.adsabs.harvard.edu/abs/2013ATel.4807....1K} {4807, 1}

\bibitem[\protect\citeauthoryear{{Kurucz}}{{Kurucz}}{2005}]{Kurucz2005}
{Kurucz} R.~L.,  2005, Memorie della Societa Astronomica Italiana Supplementi,
  \href {https://ui.adsabs.harvard.edu/abs/2005MSAIS...8...14K} {8, 14}

\bibitem[\protect\citeauthoryear{{Lachowicz}, {Zdziarski},
  {Schwarzenberg-Czerny}, {Pooley}  \& {Kitamoto}}{{Lachowicz}
  et~al.}{2006}]{Lachowicz2006}
{Lachowicz} P.,  {Zdziarski} A.~A.,  {Schwarzenberg-Czerny} A.,  {Pooley}
  G.~G.,   {Kitamoto} S.,  2006, \mn@doi [\mnras]
  {10.1111/j.1365-2966.2006.10219.x}, \href
  {https://ui.adsabs.harvard.edu/abs/2006MNRAS.368.1025L} {368, 1025}

\bibitem[\protect\citeauthoryear{{Law} et~al.,}{{Law} et~al.}{2009}]{Law2009}
{Law} N.~M.,  et~al., 2009, \mn@doi [\pasp] {10.1086/648598}, \href
  {https://ui.adsabs.harvard.edu/abs/2009PASP..121.1395L} {121, 1395}

\bibitem[\protect\citeauthoryear{{L{\'e}pine}, {Rich}  \& {Shara}}{{L{\'e}pine}
  et~al.}{2007}]{Lepine07}
{L{\'e}pine} S.,  {Rich} R.~M.,   {Shara} M.~M.,  2007, \mn@doi [\apj]
  {10.1086/521614}, \href
  {https://ui.adsabs.harvard.edu/abs/2007ApJ...669.1235L} {669, 1235}

\bibitem[\protect\citeauthoryear{{Liebert}, {Schmidt}, {Lesser}, {Stepanian},
  {Lipovetsky}, {Chaffe}, {Foltz}  \& {Bergeron}}{{Liebert}
  et~al.}{1994}]{Liebert1994}
{Liebert} J.,  {Schmidt} G.~D.,  {Lesser} M.,  {Stepanian} J.~A.,  {Lipovetsky}
  V.~A.,  {Chaffe} F.~H.,  {Foltz} C.~B.,   {Bergeron} P.,  1994, \mn@doi
  [\apj] {10.1086/173685}, \href
  {https://ui.adsabs.harvard.edu/abs/1994ApJ...421..733L} {421, 733}

\bibitem[\protect\citeauthoryear{{Lomb}}{{Lomb}}{1976}]{Lomb1976}
{Lomb} N.~R.,  1976, \mn@doi [\apss] {10.1007/BF00648343}, \href
  {https://ui.adsabs.harvard.edu/abs/1976Ap&SS..39..447L} {39, 447}

\bibitem[\protect\citeauthoryear{{Lowrance}, {Kirkpatrick}, {Reid}, {Cruz}  \&
  {Liebert}}{{Lowrance} et~al.}{2003}]{Lowrance03}
{Lowrance} P.~J.,  {Kirkpatrick} J.~D.,  {Reid} I.~N.,  {Cruz} K.~L.,
  {Liebert} J.,  2003, \mn@doi [\apjl] {10.1086/373986}, \href
  {https://ui.adsabs.harvard.edu/abs/2003ApJ...584L..95L} {584, L95}

\bibitem[\protect\citeauthoryear{{Lucatello}, {Gratton}, {Cohen}, {Beers},
  {Christlieb}, {Carretta}  \& {Ram{\'\i}rez}}{{Lucatello}
  et~al.}{2003}]{Luc03}
{Lucatello} S.,  {Gratton} R.,  {Cohen} J.~G.,  {Beers} T.~C.,  {Christlieb}
  N.,  {Carretta} E.,   {Ram{\'\i}rez} S.,  2003, \mn@doi [\aj]
  {10.1086/345886}, \href
  {https://ui.adsabs.harvard.edu/abs/2003AJ....125..875L} {125, 875}

\bibitem[\protect\citeauthoryear{{Lucatello}, {Tsangarides}, {Beers},
  {Carretta}, {Gratton}  \& {Ryan}}{{Lucatello} et~al.}{2005}]{Cemps}
{Lucatello} S.,  {Tsangarides} S.,  {Beers} T.~C.,  {Carretta} E.,  {Gratton}
  R.~G.,   {Ryan} S.~G.,  2005, \mn@doi [\apj] {10.1086/428104}, \href
  {https://ui.adsabs.harvard.edu/abs/2005ApJ...625..825L} {625, 825}

\bibitem[\protect\citeauthoryear{{Maeder}}{{Maeder}}{1987}]{Blue_S}
{Maeder} A.,  1987, \aap, \href
  {https://ui.adsabs.harvard.edu/abs/1987A&A...178..159M} {178, 159}

\bibitem[\protect\citeauthoryear{{Margon}, {Kupfer}, {Burdge}, {Prince},
  {Kulkarni}  \& {Shupe}}{{Margon} et~al.}{2018}]{Margon18}
{Margon} B.,  {Kupfer} T.,  {Burdge} K.,  {Prince} T.~A.,  {Kulkarni} S.~R.,
  {Shupe} D.~L.,  2018, \mn@doi [\apjl] {10.3847/2041-8213/aab42a}, \href
  {https://ui.adsabs.harvard.edu/abs/2018ApJ...856L...2M} {856, L2}

\bibitem[\protect\citeauthoryear{{Masci} et~al.,}{{Masci}
  et~al.}{2019}]{Masci2019}
{Masci} F.~J.,  et~al., 2019, \mn@doi [\pasp] {10.1088/1538-3873/aae8ac}, \href
  {https://ui.adsabs.harvard.edu/abs/2019PASP..131a8003M} {131, 018003}

\bibitem[\protect\citeauthoryear{{Matrozis}, {Abate}  \&
  {Stancliffe}}{{Matrozis} et~al.}{2017}]{Mat17}
{Matrozis} E.,  {Abate} C.,   {Stancliffe} R.~J.,  2017, \mn@doi [\aap]
  {10.1051/0004-6361/201730746}, \href
  {https://ui.adsabs.harvard.edu/abs/2017A&A...606A.137M} {606, A137}

\bibitem[\protect\citeauthoryear{{McClure}, {Fletcher}  \& {Nemec}}{{McClure}
  et~al.}{1980}]{Ba}
{McClure} R.~D.,  {Fletcher} J.~M.,   {Nemec} J.~M.,  1980, \mn@doi [\apjl]
  {10.1086/183252}, \href
  {https://ui.adsabs.harvard.edu/abs/1980ApJ...238L..35M} {238, L35}

\bibitem[\protect\citeauthoryear{McKinney et~al.}{McKinney
  et~al.}{2010}]{pandas}
McKinney W.,  et~al., 2010, in Proceedings of the 9th Python in Science
  Conference. pp 51--56

\bibitem[\protect\citeauthoryear{{McQuillan}, {Aigrain}  \&
  {Mazeh}}{{McQuillan} et~al.}{2013}]{McQuil}
{McQuillan} A.,  {Aigrain} S.,   {Mazeh} T.,  2013, \mn@doi [\mnras]
  {10.1093/mnras/stt536}, \href
  {https://ui.adsabs.harvard.edu/abs/2013MNRAS.432.1203M} {432, 1203}

\bibitem[\protect\citeauthoryear{{Miszalski}, {Boffin}  \&
  {Corradi}}{{Miszalski} et~al.}{2013}]{Miszalski13}
{Miszalski} B.,  {Boffin} H. M.~J.,   {Corradi} R. L.~M.,  2013, \mn@doi
  [\mnras] {10.1093/mnrasl/sls011}, \href
  {https://ui.adsabs.harvard.edu/abs/2013MNRAS.428L..39M} {428, L39}

\bibitem[\protect\citeauthoryear{{Mould}, {Schneider}, {Gordon}, {Aaronson}  \&
  {Liebert}}{{Mould} et~al.}{1985}]{Mould85}
{Mould} J.~R.,  {Schneider} D.~P.,  {Gordon} G.~A.,  {Aaronson} M.,   {Liebert}
  J.~W.,  1985, \mn@doi [\pasp] {10.1086/131508}, \href
  {https://ui.adsabs.harvard.edu/abs/1985PASP...97..130M} {97, 130}

\bibitem[\protect\citeauthoryear{{Nebot G{\'o}mez-Mor{\'a}n} et~al.,}{{Nebot
  G{\'o}mez-Mor{\'a}n} et~al.}{2011}]{Nebot2011}
{Nebot G{\'o}mez-Mor{\'a}n} A.,  et~al., 2011, \mn@doi [\aap]
  {10.1051/0004-6361/201117514}, \href
  {https://ui.adsabs.harvard.edu/abs/2011A&A...536A..43N} {536, A43}

\bibitem[\protect\citeauthoryear{{Ofek} et~al.,}{{Ofek}
  et~al.}{2012}]{Ofek2012}
{Ofek} E.~O.,  et~al., 2012, \mn@doi [\pasp] {10.1086/664065}, \href
  {https://ui.adsabs.harvard.edu/abs/2012PASP..124...62O} {124, 62}

\bibitem[\protect\citeauthoryear{{Paczynski}}{{Paczynski}}{1976}]{Pac76}
{Paczynski} B.,  1976, in {Eggleton} P.,  {Mitton} S.,   {Whelan} J.,  eds,
  IAU Symposium Vol. 73, Structure and Evolution of Close Binary Systems. p.~75

\bibitem[\protect\citeauthoryear{{Parker}}{{Parker}}{1955}]{Parker55}
{Parker} E.~N.,  1955, \mn@doi [\apj] {10.1086/146087}, \href
  {https://ui.adsabs.harvard.edu/abs/1955ApJ...122..293P} {122, 293}

\bibitem[\protect\citeauthoryear{{Plant}, {Margon}, {Guhathakurta},
  {Cunningham}, {Toloba}  \& {Munn}}{{Plant} et~al.}{2016}]{Plant16}
{Plant} K.~A.,  {Margon} B.,  {Guhathakurta} P.,  {Cunningham} E.~C.,  {Toloba}
  E.,   {Munn} J.~A.,  2016, \mn@doi [\apj] {10.3847/1538-4357/833/2/232},
  \href {https://ui.adsabs.harvard.edu/abs/2016ApJ...833..232P} {833, 232}

\bibitem[\protect\citeauthoryear{{Plez} \& {Cohen}}{{Plez} \&
  {Cohen}}{2005}]{Plez05}
{Plez} B.,  {Cohen} J.~G.,  2005, \mn@doi [\aap] {10.1051/0004-6361:20042082},
  \href {https://ui.adsabs.harvard.edu/abs/2005A&A...434.1117P} {434, 1117}

\bibitem[\protect\citeauthoryear{{Podsiadlowski}}{{Podsiadlowski}}{2001}]{Pod01}
{Podsiadlowski} P.,  2001, in {Podsiadlowski} P.,  {Rappaport} S.,  {King}
  A.~R.,  {D'Antona} F.,   {Burderi} L.,  eds,  Astronomical Society of the
  Pacific Conference Series Vol. 229, Evolution of Binary and Multiple Star
  Systems. p.~239

\bibitem[\protect\citeauthoryear{{Price-Whelan}, {Hogg}, {Foreman-Mackey}  \&
  {Rix}}{{Price-Whelan} et~al.}{2017}]{Price2017}
{Price-Whelan} A.~M.,  {Hogg} D.~W.,  {Foreman-Mackey} D.,   {Rix} H.-W.,
  2017, \mn@doi [\apj] {10.3847/1538-4357/aa5e50}, \href
  {https://ui.adsabs.harvard.edu/abs/2017ApJ...837...20P} {837, 20}

\bibitem[\protect\citeauthoryear{{Price-Whelan} et~al.,}{{Price-Whelan}
  et~al.}{2020}]{Joker}
{Price-Whelan} A.~M.,  et~al., 2020, \mn@doi [\apj] {10.3847/1538-4357/ab8acc},
  \href {https://ui.adsabs.harvard.edu/abs/2020ApJ...895....2P} {895, 2}

\bibitem[\protect\citeauthoryear{{Rau} et~al.,}{{Rau} et~al.}{2009}]{Rau2009}
{Rau} A.,  et~al., 2009, \mn@doi [\pasp] {10.1086/605911}, \href
  {https://ui.adsabs.harvard.edu/abs/2009PASP..121.1334R} {121, 1334}

\bibitem[\protect\citeauthoryear{{Rebassa-Mansergas}, {Schreiber}  \&
  {G{\"a}nsicke}}{{Rebassa-Mansergas} et~al.}{2013}]{Rebassa13}
{Rebassa-Mansergas} A.,  {Schreiber} M.~R.,   {G{\"a}nsicke} B.~T.,  2013,
  \mn@doi [\mnras] {10.1093/mnras/sts630}, \href
  {https://ui.adsabs.harvard.edu/abs/2013MNRAS.429.3570R} {429, 3570}

\bibitem[\protect\citeauthoryear{{Reid} \& {Gizis}}{{Reid} \&
  {Gizis}}{2005}]{Reid05}
{Reid} I.~N.,  {Gizis} J.~E.,  2005, \mn@doi [\pasp] {10.1086/430462}, \href
  {https://ui.adsabs.harvard.edu/abs/2005PASP..117..676R} {117, 676}

\bibitem[\protect\citeauthoryear{{Ricker} et~al.,}{{Ricker}
  et~al.}{2015}]{TESS}
{Ricker} G.~R.,  et~al., 2015, \mn@doi [Journal of Astronomical Telescopes,
  Instruments, and Systems] {10.1117/1.JATIS.1.1.014003}, \href
  {https://ui.adsabs.harvard.edu/abs/2015JATIS...1a4003R} {1, 014003}

\bibitem[\protect\citeauthoryear{{Rossi}, {Gigoyan}, {Avtandilyan}  \&
  {Sclavi}}{{Rossi} et~al.}{2011}]{Rossi2011}
{Rossi} C.,  {Gigoyan} K.~S.,  {Avtandilyan} M.~G.,   {Sclavi} S.,  2011,
  \mn@doi [\aap] {10.1051/0004-6361/201116704}, \href
  {https://ui.adsabs.harvard.edu/abs/2011A&A...532A..69R} {532, A69}

\bibitem[\protect\citeauthoryear{{Roulston} et~al.,}{{Roulston}
  et~al.}{2019}]{Roulston19}
{Roulston} B.~R.,  et~al., 2019, \mn@doi [\apj] {10.3847/1538-4357/ab1a3e},
  \href {https://ui.adsabs.harvard.edu/abs/2019ApJ...877...44R} {877, 44}

\bibitem[\protect\citeauthoryear{{Roulston}, {Green}, {Toonen}  \&
  {Hermes}}{{Roulston} et~al.}{2021}]{Roulston2021}
{Roulston} B.~R.,  {Green} P.~J.,  {Toonen} S.,   {Hermes} J.~J.,  2021, arXiv
  e-prints, \href {https://ui.adsabs.harvard.edu/abs/2021arXiv210500037R} {p.
  arXiv:2105.00037}

\bibitem[\protect\citeauthoryear{Salvatier, Wiecki  \& Fonnesbeck}{Salvatier
  et~al.}{2016}]{pymc3}
Salvatier J.,  Wiecki T.~V.,   Fonnesbeck C.,  2016, \mn@doi [{PeerJ} Computer
  Science] {10.7717/peerj-cs.55}, 2, e55

\bibitem[\protect\citeauthoryear{{Scargle}}{{Scargle}}{1982}]{Scargle1982}
{Scargle} J.~D.,  1982, \mn@doi [\apj] {10.1086/160554}, \href
  {https://ui.adsabs.harvard.edu/abs/1982ApJ...263..835S} {263, 835}

\bibitem[\protect\citeauthoryear{{Schwarzenberg-Czerny}}{{Schwarzenberg-Czerny}}{1996}]{Schwarz1996}
{Schwarzenberg-Czerny} A.,  1996, \mn@doi [\apjl] {10.1086/309985}, \href
  {https://ui.adsabs.harvard.edu/abs/1996ApJ...460L.107S} {460, L107}

\bibitem[\protect\citeauthoryear{{Skumanich}}{{Skumanich}}{1972}]{Skum}
{Skumanich} A.,  1972, \mn@doi [\apj] {10.1086/151310}, \href
  {https://ui.adsabs.harvard.edu/abs/1972ApJ...171..565S} {171, 565}

\bibitem[\protect\citeauthoryear{{Temmink}, {Toonen}, {Zapartas}, {Justham}  \&
  {G{\"a}nsicke}}{{Temmink} et~al.}{2020}]{Merg}
{Temmink} K.~D.,  {Toonen} S.,  {Zapartas} E.,  {Justham} S.,   {G{\"a}nsicke}
  B.~T.,  2020, \mn@doi [\aap] {10.1051/0004-6361/201936889}, \href
  {https://ui.adsabs.harvard.edu/abs/2020A&A...636A..31T} {636, A31}

\bibitem[\protect\citeauthoryear{{Theuns}, {Boffin}  \& {Jorissen}}{{Theuns}
  et~al.}{1996}]{Theuns96}
{Theuns} T.,  {Boffin} H. M.~J.,   {Jorissen} A.,  1996, \mn@doi [\mnras]
  {10.1093/mnras/280.4.1264}, \href
  {https://ui.adsabs.harvard.edu/abs/1996MNRAS.280.1264T} {280, 1264}

\bibitem[\protect\citeauthoryear{{Tody}}{{Tody}}{1993}]{Iraf}
{Tody} D.,  1993, in {Hanisch} R.~J.,  {Brissenden} R.~J.~V.,   {Barnes} J.,
  eds,  Astronomical Society of the Pacific Conference Series Vol. 52,
  Astronomical Data Analysis Software and Systems II. p.~173

\bibitem[\protect\citeauthoryear{{Tonry} \& {Davis}}{{Tonry} \&
  {Davis}}{1979}]{TD1979}
{Tonry} J.,  {Davis} M.,  1979, \mn@doi [\aj] {10.1086/112569}, \href
  {https://ui.adsabs.harvard.edu/abs/1979AJ.....84.1511T} {84, 1511}

\bibitem[\protect\citeauthoryear{{Van Eck} et~al.,}{{Van Eck}
  et~al.}{2017}]{VEck17}
{Van Eck} S.,  et~al., 2017, \mn@doi [\aap] {10.1051/0004-6361/201525886},
  \href {https://ui.adsabs.harvard.edu/abs/2017A&A...601A..10V} {601, A10}

\bibitem[\protect\citeauthoryear{{VanderPlas}}{{VanderPlas}}{2018}]{Van2018}
{VanderPlas} J.~T.,  2018, \mn@doi [\apjs] {10.3847/1538-4365/aab766}, \href
  {https://ui.adsabs.harvard.edu/abs/2018ApJS..236...16V} {236, 16}

\bibitem[\protect\citeauthoryear{Virtanen et~al.,}{Virtanen
  et~al.}{2020}]{Scipy}
Virtanen P.,  et~al., 2020, \mn@doi [Nature Methods]
  {10.1038/s41592-019-0686-2}, \href {https://rdcu.be/b08Wh} {17, 261}

\bibitem[\protect\citeauthoryear{{Webbink}}{{Webbink}}{1984}]{Webbink84}
{Webbink} R.~F.,  1984, \mn@doi [\apj] {10.1086/161701}, \href
  {https://ui.adsabs.harvard.edu/abs/1984ApJ...277..355W} {277, 355}

\bibitem[\protect\citeauthoryear{{Whitehouse}, {Farihi}, {Green}, {Wilson}  \&
  {Subasavage}}{{Whitehouse} et~al.}{2018}]{Whitehouse18}
{Whitehouse} L.~J.,  {Farihi} J.,  {Green} P.~J.,  {Wilson} T.~G.,
  {Subasavage} J.~P.,  2018, \mn@doi [\mnras] {10.1093/mnras/sty1622}, \href
  {https://ui.adsabs.harvard.edu/abs/2018MNRAS.479.3873W} {479, 3873}

\bibitem[\protect\citeauthoryear{{Zahn}}{{Zahn}}{1989}]{Zahn89}
{Zahn} J.~P.,  1989, \aap, \href
  {https://ui.adsabs.harvard.edu/abs/1989A&A...220..112Z} {220, 112}

\bibitem[\protect\citeauthoryear{{Zahn}}{{Zahn}}{2008}]{Zahn}
{Zahn} J.~P.,  2008, in {Goupil} M.~J.,  {Zahn} J.~P.,  eds,  EAS Publications
  Series Vol. 29, EAS Publications Series. pp 67--90 (\mn@eprint {arXiv}
  {0807.4870}), \mn@doi{10.1051/eas:0829002}

\bibitem[\protect\citeauthoryear{{Zechmeister} \& {K{\"u}rster}}{{Zechmeister}
  \& {K{\"u}rster}}{2009}]{Zech09}
{Zechmeister} M.,  {K{\"u}rster} M.,  2009, \mn@doi [\aap]
  {10.1051/0004-6361:200811296}, \href
  {https://ui.adsabs.harvard.edu/abs/2009A&A...496..577Z} {496, 577}

\makeatother
\end{thebibliography}


\bigskip
\bigskip



\appendix

\section{Posterior Distributions for radial-velocity Determinations}

\begin{figure*}
	\includegraphics[width=0.495\textwidth]{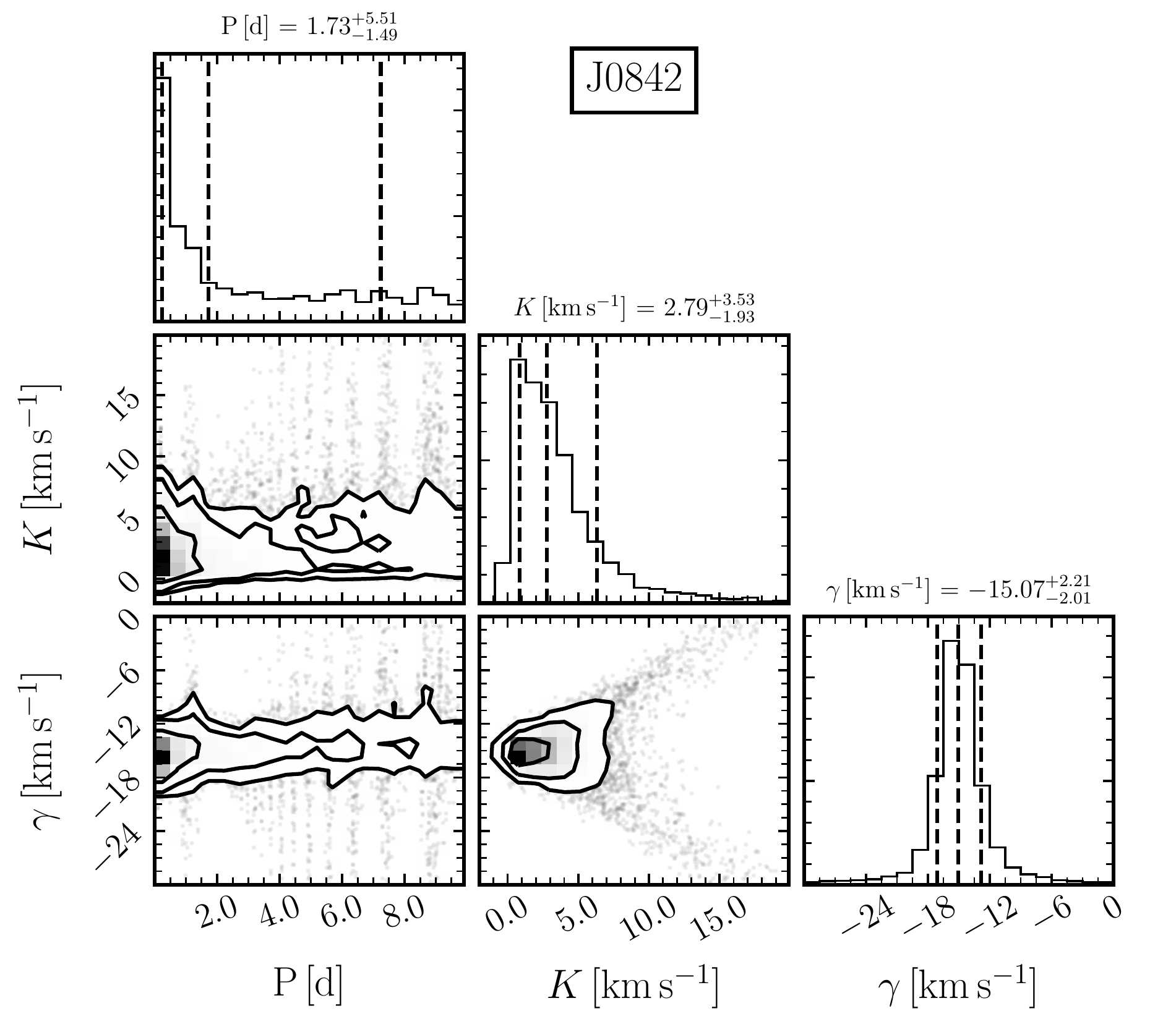}
	\includegraphics[width=0.495\textwidth]{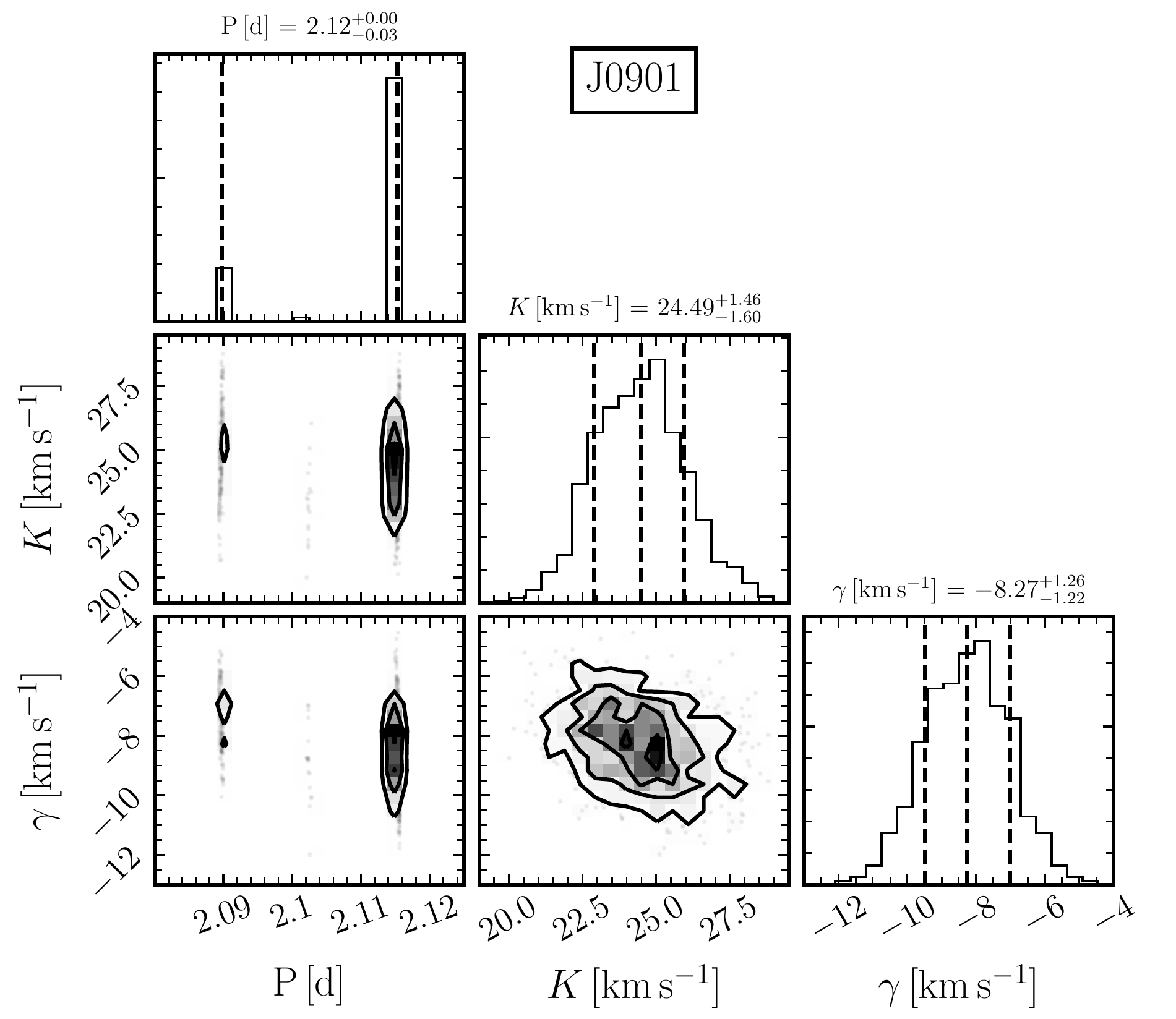}
 	\includegraphics[width=0.495\textwidth]{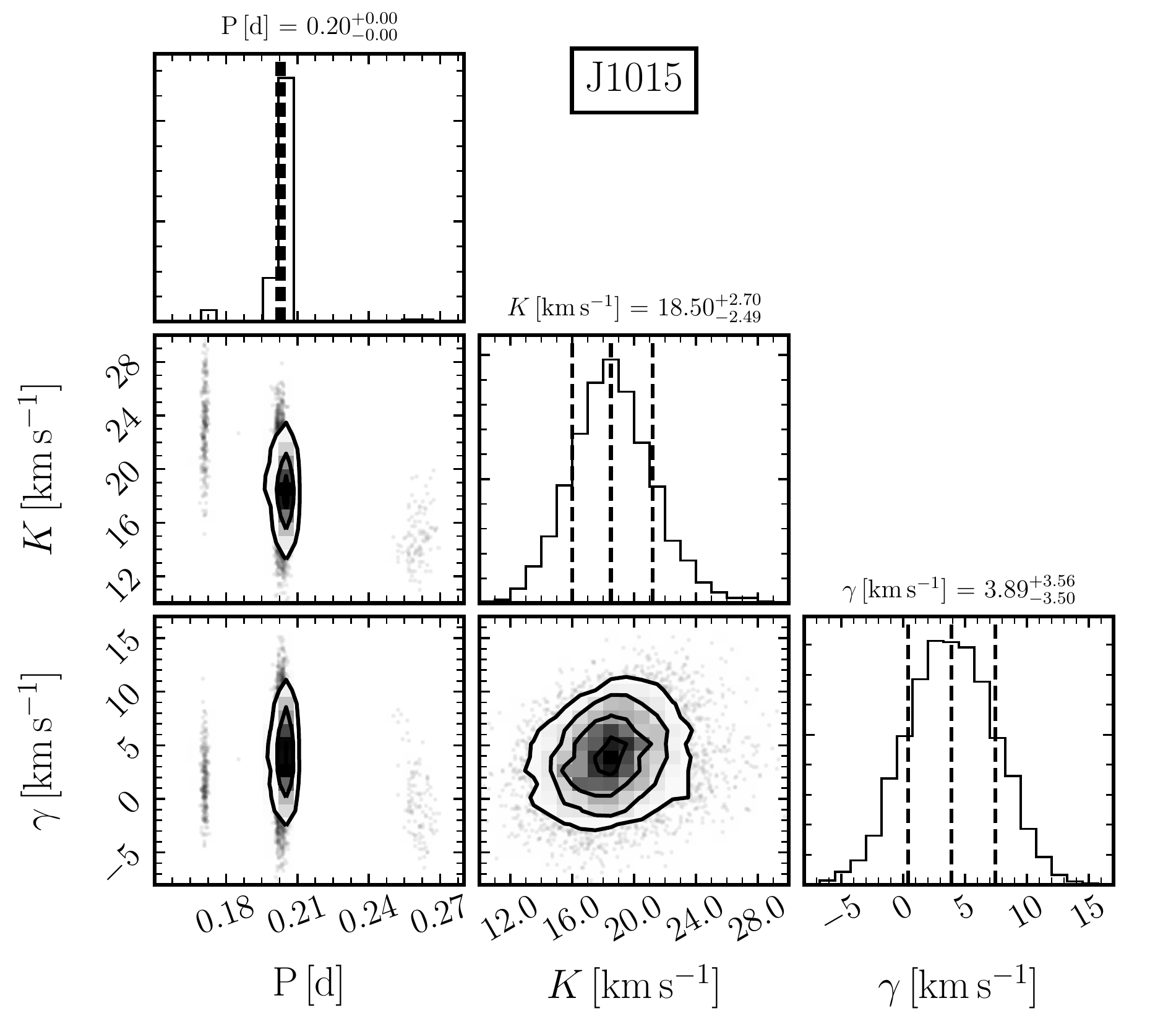}
 	\includegraphics[width=0.495\textwidth]{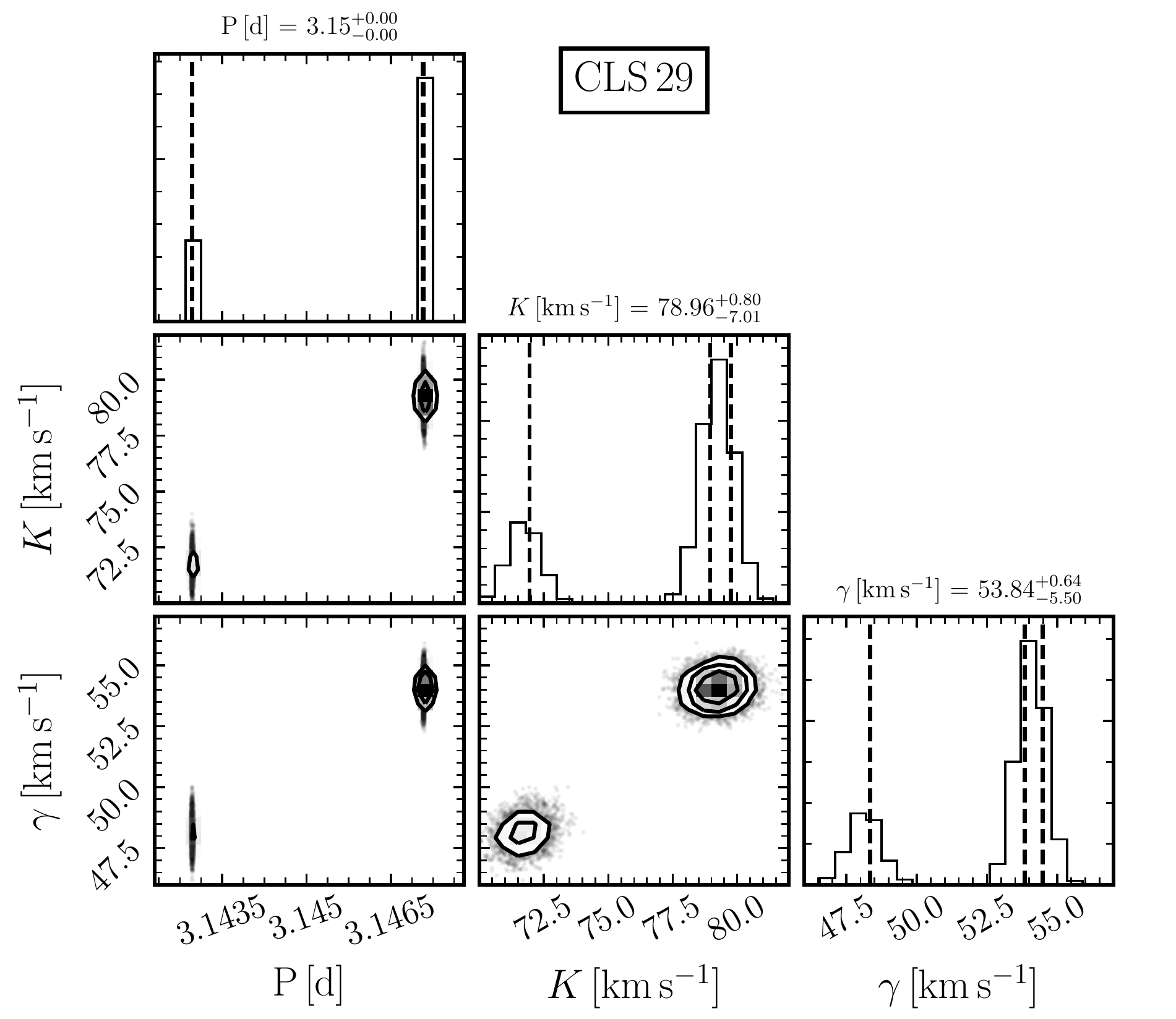}
\caption{The radial-velocity posterior PDFs for all targets analysed using the \textsc{joker} rejection sampler presented in order of right ascension; J0842, J0901, J1015, CLS\,29, J1250, SBSS\,1310, and CBS\,311.  The orbital period, velocity semi-amplitude, and systemic velocities are presented along with the 16\,per cent and 84\, per cent confidence limits for each target.}
	\label{fig:all_corner}
\end{figure*}

\begin{figure*}
	\includegraphics[width=0.495\textwidth]{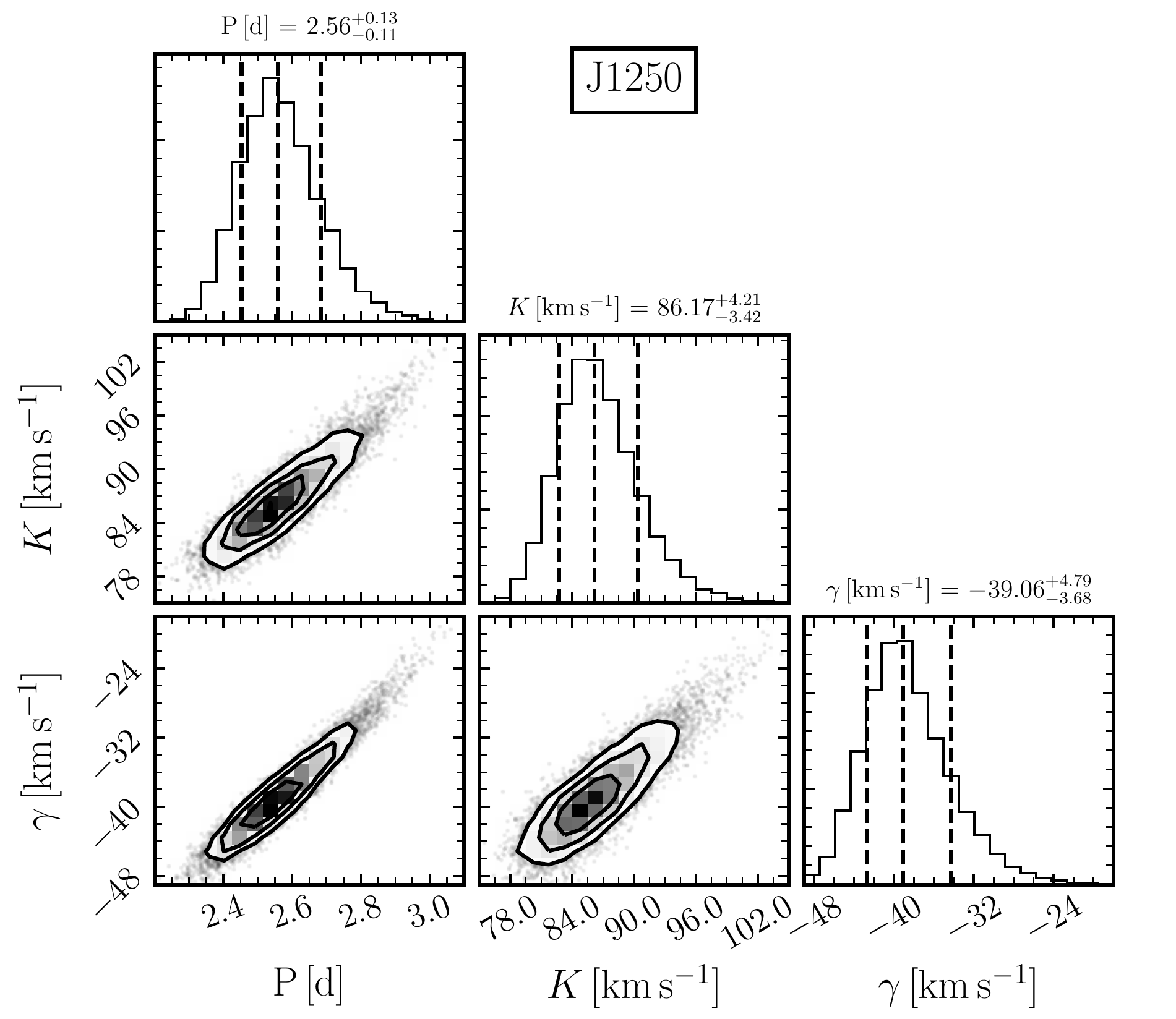}
	\includegraphics[width=0.495\textwidth]{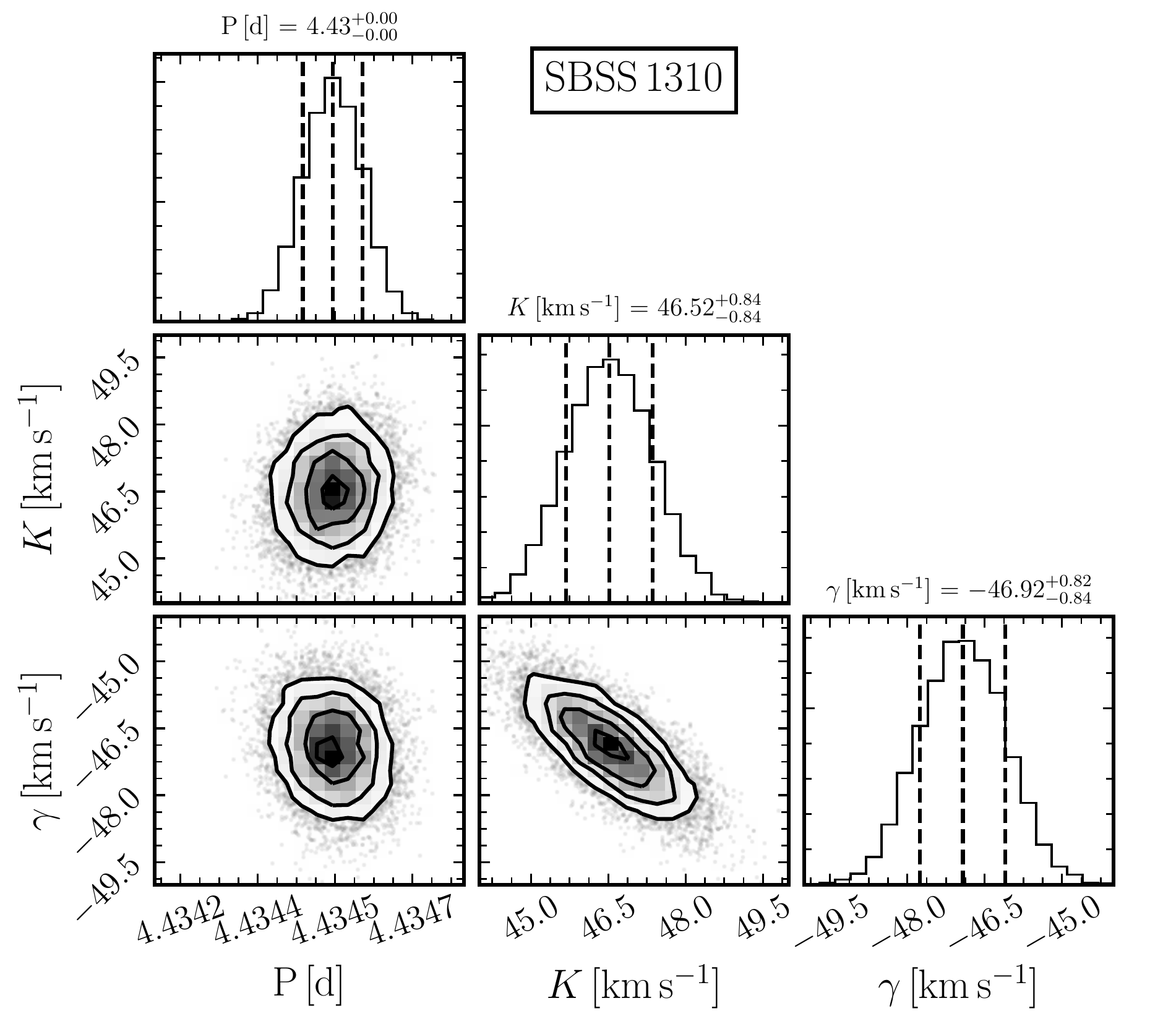}
 	\includegraphics[width=0.495\textwidth]{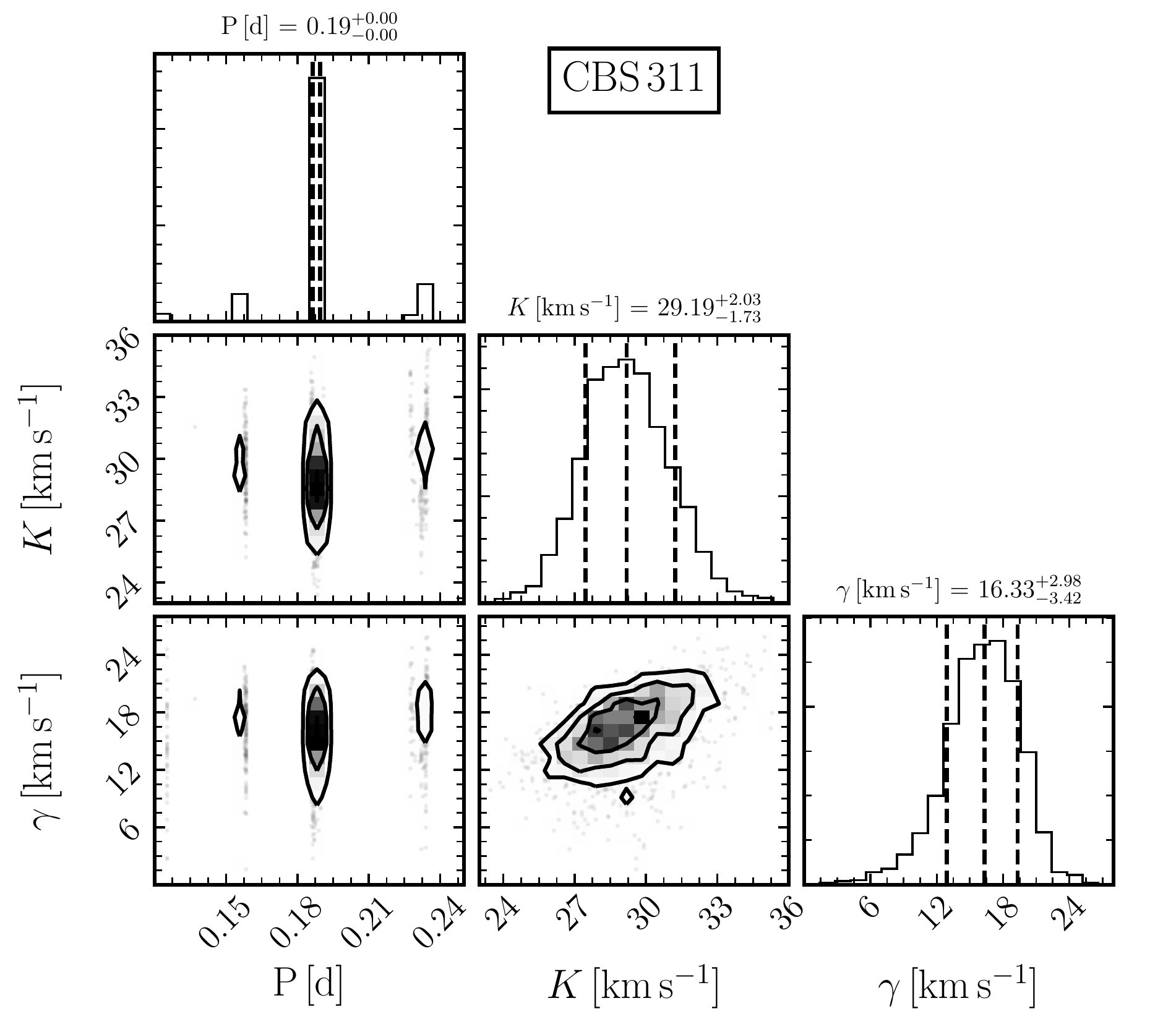}
 	\contcaption{}
\end{figure*}


\bsp	
\label{lastpage}
\end{document}